\documentclass[epsf,pra,twocolumn,showpacs,nofootinbib,longbibliography]{revtex4-1}
\usepackage{amsmath,amssymb,amsthm,amscd,latexsym,epsfig,bm}
\usepackage{float}
\usepackage[justification=raggedright,font=small]{caption}
\usepackage{graphicx}
\graphicspath{{images/}}
\usepackage{subcaption}
\usepackage{color}
\usepackage{framed}
\usepackage[colorlinks,bookmarks=false,citecolor=blue,linkcolor=red,urlcolor=blue]{hyperref}
\usepackage{verbatim}
\usepackage[normalem]{ulem}

\usepackage[dvipsnames]{xcolor}

\newcommand{\zz}{\mathbb{Z}_2}
\newcommand{\z}{\mathbb{Z}}

\def\hx{\hat{{x}}}
\def\hy{\hat{{y}}}
\def\hz{\hat{{z}}}

\def\br{\boldsymbol{r}}
\def\bR{\boldsymbol{R}}

\def\ba{\boldsymbol{a}}
\def\zt{{\mathbb Z}[T]}

\begin{document}

\title{Fracton fusion and statistics}
\author{Shriya Pai}
\author{Michael Hermele}
\affiliation{Department of Physics and Center for Theory of Quantum Matter, University of Colorado, Boulder, CO 80309, USA}
\date{\today}

\begin{abstract}
We introduce and develop a theory of fusion and statistical processes of gapped excitations in Abelian fracton phases.  The key idea is to incorporate lattice translation symmetry via its action on superselection sectors, which results in a fusion theory endowed with information about the non-trivial mobility of fractons and sub-dimensional excitations.  This results in a description of statistical processes in terms of local moves determined by the fusion theory.  Our results can be understood as providing a characterization of translation-invariant fracton phases. We obtain simple descriptions of the fusion theory in the X-cube and checkerboard fracton models, as well as for gapped electric and magnetic excitations of some gapless ${\rm U}(1)$ tensor gauge theories.  An alternate route to the X-cube model fusion theory is provided by starting with a system of decoupled two-dimensional toric code layers, and giving a description of the $p$-string condensation mechanism within our approach. We discuss examples of statistical processes of fractons and sub-dimensional excitations in the X-cube and checkerboard models.  As an application of the ideas developed, we prove that the X-cube and semionic X-cube models realize distinct translation-invariant fracton phases, even when the translation symmetry is broken corresponding to an arbitrary but finite enlargement of the crystalline unit cell.
\end{abstract}

\maketitle

\tableofcontents

\section{Introduction}
\label{sec:intro}

Quantum many-body systems with fracton  excitations have recently emerged as a frontier of condensed matter physics.  Following the early work of Chamon \cite{chamon}, Haah \cite{haah11local} discovered a class of fracton states in a quest to realize a stable finite-temperature quantum memory.  More recently, spurred on by the development of simpler solvable fracton models \cite{fracton1,vijay16fracton} and by the observation that many ${\rm U}(1)$ symmetric-tensor gauge theories \cite{Xu1, Pankov, XuWuPlaquette,  Xu2}  support fracton excitations \cite{pretko17subdimensional}, interest has grown in fracton states as a new class of quantum phases of matter that lie beyond existing paradigms.  The key property is the presence of excitations of restricted mobility, including sub-dimensional particles that move along lines (``lineons'') or planes (``planons''), and fractons that, individually, cannot move at all.  For a recent brief review of fracton phases, see Ref.~\onlinecite{fractonarcmp}.

There are both gapped and gapless fracton phases in three spatial dimensions.  In many respects, gapped fracton phases are similar to more conventional phases with intrinsic topological order (iTO) \cite{wen1989,wen1990,wen1990groundstate,wen2013}:  there is an energy gap to all excitations, and some excitations above the gap are topologically non-trivial in the sense that individual such excitations cannot be created locally.  Moreover, on the torus there is a degeneracy among ground states that cannot be distinguished by local measurements.  However, in gapped fracton phases, some of the topologically non-trivial excitations are fractons or sub-dimensional particles, and the ground state degeneracy on the torus grows sub-extensively with system size \cite{bravyi11topological,haah11local,vijay16fracton}.  In the gapless fracton phases that are currently best-understood,  fractons and sub-dimensional particles arise as gapped electric and magnetic excitations of a ${\rm U}(1)$ tensor gauge theory, which also has gapless photon-like excitations \cite{pretko17subdimensional}.

While by now there are many models exhibiting fracton phases, there is still relatively little understanding of the theoretical  characterization of these phases (but see below for a brief survey of prior such work).  By a characterization we mean a set of properties, and perhaps a theoretical structure encapsulating these properties, that are both in principle measurable in experiments or numerical simulations, and that are universal in the sense that they provide robust distinctions between different phases of matter. One important way to characterize conventional iTO phases in two dimensions is in terms of the topologically non-trivial gapped quasiparticle excitations \cite{halperin84statistics, arovas84fractional,wen91nonAbelian,moore91nonAbelions,kitaev2006anyons,wen15theory}.  At the most basic level, one enumerates the distinct types of excitations that exist above a given ground state.  Next, one studies fusion of excitations; namely, we can ask what type of excitation arises when we make a composite of two excitations of definite type.  Finally, one can consider statistical processes where excitations are braided around one another.  Such characterizations are important not only in fully gapped topologically ordered states, but also for gapped electric and magnetic excitations of (ordinary 1-form) ${\rm U}(1)$ gauge theories in three spatial dimensions.

In this paper, we follow a similar path and describe the characterization of fracton phases in terms of their gapped excitations.  In particular, we describe fusion and statistical processes in Abelian fracton phases, by which we mean those fracton phases where fusing two excitations gives a composite excitation of definite type (see Sec.~\ref{subsec:fusion-preliminaries}). The focus throughout is on three spatial dimensions and on type I fracton phases \cite{vijay16fracton}, and in particular those type I phases without fractal structure, \emph{e.g.} where fractons are created at the corners of membrane operators rather than operators with fractal support.  (We recall that type I fracton phases are defined to be those with at least one non-trivial string operator.)  We also focus on systems where the underlying degrees of freedom are bosonic; the generalization to fermionic systems is straightforward. Our approach to fusion of excitations is equally applicable to type II phases, but this is less clear for statistical processes, as we discuss briefly in Sec.~\ref{sec:discussion}. Most of the paper focuses on gapped fracton phases, including detailed discussion of the examples of the X-cube and checkerboard fracton models \cite{vijay16fracton}, and the semionic X-cube model \cite{ma17coupled}.  In addition, we discuss fusion of gapped electric and magnetic excitations in symmetric-tensor gauge theories, using two members of the family of $(m,n)$ scalar charge theories \cite{bulmash18Higgs} as illustrative examples.

As compared to more conventional states, there are two key new features in fracton phases.  First is the restricted mobility of excitations, which implies that statistical processes need not always take the form of familiar braiding processes.  The fusion theory we develop encodes the mobility of excitations, which allows us to use it as a starting point to describe statistical processes.  Second, the number of distinct excitation types in fracton phases in infinite, in contrast to conventional iTO phases.  This strongly suggests that, in order to get a manageable theory, we need to impose some structure beyond what is present in the theory of conventional iTO phases.

To build a theory that incorporates these features, we consider lattice translation symmetry.  If we ignore translation symmetry, the fusion of excitations in an Abelian fracton phase is described by an infinite Abelian group, whose elements correspond to distinct excitation types.  Translation symmetry acts on this Abelian group, giving it more structure and making it into a more manageable object.  Moreover, this action directly allows us to describe the mobility of excitations at the level of the fusion theory, which then forms the basis for a description of statistical processes.

From a certain point of view, it may appear undesirable to incorporate translation symmetry; for instance, gapped fracton phases are robust in the absence of any symmetries.  However, it is by now a common observation that discrete lattice geometry seems to play a key role in the physics of fracton phases, and it appears that some geometrical structure may be a necessary ingredient of a tractable theory.  An important example is the theory of foliated fracton phases \cite{shirley18manifolds,shirley19entanglement,shirley2019fractional,shirley18checkerboard,shirley18subsystem,slagle18foliated}, described in more detail below, where geometrical information is provided by a certain kind of foliation structure. Lattice translation imposes a different kind of geometrical structure complementary to that provided by foliation.  In addition, our approach is still useful if the translation symmetry is lowered to a subgroup, \emph{i.e.} if the crystalline unit cell is enlarged, and can be used to study distinctions between phases that are preserved or lost upon such lowering of symmetry.  A number of other works have also investigated the interplay between lattice geometry or symmetry and fracton order, from a variety of perspectives \cite{haah13commuting, haah16algebraic,slagle17robust,slagle17generic,pretko18elasticity,pretko18supersolid,kumar18fractonicity,williamson18SETfracton,gromov18multipole,dua2019compactification}.

Another perspective on fracton phases emphasizes the importance of emergent conservation laws that lead to the restricted mobility of excitations \cite{pretko17subdimensional}.  One important example is the conservation of electric charge and dipole moment in the so-called rank-2 scalar charge ${\rm U}(1)$ tensor gauge theory; conservation of dipole moment makes isolated electric charges into immobile fractons \cite{pretko17subdimensional}.  Another is the conservation of the number of cube excitations modulo two on each lattice plane in the X-cube model \cite{vijay16fracton}.  This point of view is also present in conventional iTO phases, and is encoded in the fusion theory.  For instance, the $e$-particles (vertex excitations) and $m$-particles (plaquette excitations) in the $d=2$ toric code \cite{kitaev2003fault} cannot be created individually but only in pairs.  This can be understood in terms of conservation of $\zz \oplus \zz$ ``topological charge,'' which is really just another way of saying that fusion of excitations is described by the Abelian group $\zz \oplus \zz$.  It is thus not surprising that the fusion theories we develop for Abelian fracton phases can also be understood as a way of encoding the emergent conservation laws.  This perspective plays an important role in our work, in part by guiding our computations of fusion theories in specific models.

Before proceeding to an outline of the paper, we now briefly survey some prior work on characterizations of Abelian fracton phases, including a discussion of foliated fracton phases. (Some work has also studied fracton phases with non-Abelian excitations \cite{vijay17nonabelian,song18twisted,prem18cagenet, bulmash2019gauging,prem2019gauging}.)  Gapless fracton phases can of course be partly characterized by signatures such as spin-spin correlations and heat capacity \cite{prem18pinchpoint}.  In gapped fracton phases, one important characterization tool has been ground state degeneracy \cite{bravyi11topological, haah11local, vijay16fracton,slagle17robust}, but as in conventional iTO phases, this property depends on boundary conditions that have to be specified.  Other works have studied fracton phases using entanglement entropy \cite{shi18deciphering, ma18entanglement,he18entanglement,shirley19entanglement}, in terms of nearby phases that are accessed when certain excitations are condensed \cite{bulmash18Higgs,ma18Higgs}, and in terms of generalizations of Wilson loop observables \cite{devakul18correlation}. Many works have studied gapped fracton and sub-dimensional excitations, including discussions of statistical properties and remote detection \cite{vijay16fracton,ma17coupled, slagle17robust, yizhi18twisted,shirley2019fractional, song18twisted,prem18cagenet, schmitz18gauge, bulmash18braiding}. Refs.~\onlinecite{song18twisted,prem18cagenet} introduced the notion of intrinsically sub-dimensional excitations, which is closely related to the notion of quotient superselection sectors introduced in~\onlinecite{shirley2019fractional}.  Finally, closest to our approach, in the context of stabilizer codes, Ref.~\onlinecite{haah13commuting} introduced the same mathematical object that constitutes our fusion theory.  Later, Ref.~\onlinecite{haah16algebraic} pointed out that this object can be used to characterize stabilizer codes, but did not consider statistical properties.

Perhaps the best developed characterizations of gapped fracton phases are those based on the notion of foliated fracton phases \cite{shirley18manifolds,shirley19entanglement,shirley2019fractional,shirley18checkerboard,shirley18subsystem,slagle18foliated}.  Some fracton models can be defined on general three-dimensional spatial manifolds with a certain kind of foliation structure.  Upon choosing a foliation structure, two states are considered to be in the same foliated fracton phase if they are adiabatically connected to one another, possibly after stacking with layers of $d=2$ topologically ordered states placed on the ``leaves'' of the foliation.  With this notion of equivalence, a stack of $d=2$ topologically ordered states is considered to be trivial.
This viewpoint led to an understanding of the X-cube fracton phase as a kind of renormalization group fixed point \cite{shirley18manifolds}, and also led to entanglement signatures \cite{shirley19entanglement} and characterizations of gapped excitations \cite{shirley2019fractional} that are only sensitive to which foliated phase a system is in.

The equivalence relation used to define foliated fracton phases is different from that conventionally used to define quantum phases of matter.  Usually, two states are said to be in the same phase if they can be adiabatically connected to one another, possibly after adding degrees of freedom in a trivial product state.  Adding such trivial degrees of freedom is not an arbitrary rule, but comes from the fact that lattice models are idealizations where some degrees of freedom (\emph{e.g.} core levels) are ignored, and we want the notion of a phase of matter to be independent of whether we ignore or include such ``inert'' degrees of freedom.  In defining foliated fracton phases, one allows for adding a wider range of degrees of freedom, and the resulting equivalence relation is coarser than the more conventional one.  In this paper, our discussion of translation-invariant fracton phases is based on the usual equivalence relation used to define quantum phases of matter, with translation symmetry imposed as an extra condition.  So for instance a stack of $d=2$ topological orders is a trivial foliated fracton phase but a non-trivial translation-invariant fracton phase. A more interesting example is provided by the X-cube fracton model \cite{vijay16fracton}, and its semionic variant \cite{ma17coupled}, which were shown to be in the same foliated fracton phase.  In Sec.~\ref{sec:2Xcube} we prove that these models realize distinct translation-invariant fracton phases, even when translation symmetry is broken corresponding to an arbitrary but finite enlargement of the crystalline unit cell.  In Sec.~\ref{sec:discussion}, we comment further on these two different notions of fracton phases.

\subsection{Outline}
\label{subsec:outline}

We now provide an outline of the remainder of the paper.  In Sec.~\ref{sec:fusion}, we develop the theory of fusion of excitations in gapped Abelian fracton phases.  In Sec.~\ref{subsec:fusion-preliminaries}, we describe the fusion theory as an Abelian group, and then in Sec.~\ref{subsec:fusion-translation} we include the action of lattice translation symmetry, which makes the fusion theory into a $\zt$-module, where $\zt$ is the group ring of the translation group $T \simeq \z^3$, with integer coefficients.  Section~\ref{subsec:mobility} describes the simplest way in which the fusion theory allows one to study the mobility of excitations and identify fractons and sub-dimensional particles.

In Section~\ref{sec:examples}, we compute the fusion theory for the X-cube and checkerboard models, which are solvable type I fracton models introduced in Ref.~\onlinecite{vijay16fracton}.  By ``compute the fusion theory,'' we mean not only that we define it starting from the lattice model, but that we give a simple mathematical description that makes the fusion theory easy to work with.  We do this starting from the lattice model in Sec.~\ref{sec:xcube-fusion} for the X-cube model, and in Sec.~\ref{sec:checkerboard-fusion} for the checkerboard model.  Some technical details are relegated to Appendix~\ref{app:technical}.  In Sec.~\ref{sec:checkerboard-fusion}, we compute the fusion theory of the X-cube model in terms of $p$-string condensation, where the X-cube model is constructed from a stack of $d=2$ toric code layers, which are coupled together by condensing  extended objects dubbed $p$-strings \cite{ma17coupled,vijay17coupled}.  This approach allows us to easily see that the semionic X-cube model introduced in Ref.~\onlinecite{ma17coupled} has the same fusion theory as the ordinary X-cube model.

Section~\ref{sec:statistics} develops a description of statistical processes in terms of ``local moves.''  The notion of local moves is first introduced in Sec.~\ref{subsec:localmoves}, and then statistical processes are described in terms of local moves in Sec.~\ref{subsec:statproc}, using the $d=2$ toric code as a familiar illustrative example.  In Sec.~\ref{sec:statfracton} we proceed to describe examples of statistical processes in the X-cube and checkerboard fracton models.  We discuss two processes in the X-cube model, a fracton-lineon process (Sec.~\ref{subsec:fracton-lineon}), and a lineon-lineon process corresponding to exchange of two lineon excitations (Sec.~\ref{subsec:lineon-lineon}).  We also briefly discuss a fracton-fracton statistical process in the checkerboard model (Sec.~\ref{subsec:checkerboard-fracton-fracton}).  Section~\ref{sec:2Xcube} uses the fusion theory and the lineon-lineon statistical process to prove that the X-cube and semionic X-cube models realize distinct translation-invariant fracton phases.

In Sec.~\ref{sec:u1fusion}, we discuss the fusion of gapped electrically and magnetically charged excitations in symmetric ${\rm U}(1)$ tensor gauge theories.  We focus on the family of $(m,n)$ scalar charge theories on the simple cubic lattice \cite{bulmash18Higgs}, and in particular on the $(1,1)$ and $(2,1)$ members of this family.  We compute the fusion theory for both electric and magnetic charges of the $(1,1)$ theory in Sec.~\ref{sec:11scalarcharge}, and for electric charges in the $(2,1)$ theory in Sec.~\ref{sec:21scalarcharge}), with some technical details given in Appendix~\ref{app:technical}.  The electric fusion theories are different, with the $(2,1)$ theory enjoying additional conservation laws not present in the $(1,1)$ theory.  The $(1,1)$ and $(2,1)$ theories thus describe distinct translation-invariant gapless fracton phases.  The paper concludes in Sec.~\ref{sec:discussion} with a discussion of open questions and some general remarks on the two different notions of translation-invariant and foliated fracton phases.

\section{Fusion in Abelian fracton phases}
\label{sec:fusion}

\subsection{Preliminaries: fusion theory as an Abelian group}
\label{subsec:fusion-preliminaries}

We consider a quantum system in three spatial dimensions with a local Hamiltonian and a gap to all bulk excitations.  To study excitations above the ground state, we take the thermodynamic limit, and consider excitations with bounded support.  That is, we consider excitations whose local density matrices only differ from those of a ground state within a bounded region.  There is a limiting procedure implied here, because in order to create arbitrary such excitations within a bounded region, other excitations generally also need to be created somewhere far away from the bounded region.  We are taking a limit where these far-away excitations are pushed infinitely far from the bounded region.

Excitations can then be labeled by superselection sectors (also referred to as particle types or excitation types): two excitations are in the same sector if they can be transformed into one another by acting with an operator of bounded support, while two excitations are in different sectors if this is not possible.  There is a trivial superselection sector consisting of excitations that can be created from a ground state by acting with operators of bounded support.  We say these trivial excitations are locally createable, while non-trivial excitations are those that are not locally createable.  We denote the set of all superselection sectors by ${\cal S}$.

The scheme we have described is applicable to point-like excitations, but of course there are extended excitations in some gapped phases, such as flux loops in discrete gauge theories.  We will restrict attention to point-like excitations for simplicity and because many fracton phases only have such excitations.  While there are ${\rm U}(1)$ tensor gauge theories with line excitations exhibiting restricted mobility \cite{pai18fractonicline}, in the sense that the line objects are not flexible and able to deform their shape arbitrarily, it is not known whether such phenomena are possible in fully gapped fracton phases.

%\textcolor{Green}{While the existence of ${\rm U}(1)$ tensor gauge theories with line excitations exhibiting restricted mobility \cite{pai18fractonicline} have been known for a while, there have also been some recent examples of gapped fracton phases supporting both point-like and loop-like excitations \cite{bulmash2019gauging,prem2019gauging}}.

Now we discuss fusion of excitations.  We consider two excitations supported within disjoint and well-separated bounded regions $R_1$ and $R_2$, belonging to sectors $s_1$ and $s_2$, respectively.  We then consider a state containing both excitations; that is, the local density matrices in the regions $R_1$ and $R_2$ are the same as those in the corresponding single-excitation states, and local density matrices outside these regions are the same as in a ground state.  Then, we ask to which superselection sector $s$ this composite state belongs.  In general, with the information we specified, there may be more than one possibility for $s$.  However, we assume that $s$ is uniquely specified by $s_1$ and $s_2$; this is what it means for the excitations to be Abelian.  Moreover, given a pair of superselection sectors $s_1$ and $s_2$, we assume that we can always find representative excitations in each sector supported within disjoint and well-separated bounded regions $R_1$ and $R_2$; this assumption allows us to define a binary operation mapping ${\cal S} \times {\cal S} \to {\cal S}$, and can easily be shown to hold for all the fracton models considered in this paper.

On physical grounds, with the assumption that all excitations are Abelian, the binary operation given by fusion makes ${\cal S}$ into an Abelian group, and we use additive notation for the fusion operation.  Commutativity and associativity follow from the expectation that the result of fusion should not depend on any ordering of the excitations.  Each excitation must have an anti-particle excitation, so that the composite of a particle and its anti-particle can be created locally; therefore each sector has an inverse.

We note that there can be additional structure tied to fusion associativity, beyond the Abelian group structure.  If we look at the effect of the fusion operation $(s_1 + s_2)  + s_3$ versus $s_1 + (s_2 + s_3)$ on the vector space of states, rather than merely looking at the resulting superselection sector, the two operations can differ by a phase factor encoded in an $F$ symbol.  It is well known that some theories of Abelian anyons in two dimensions have non-trivial $F$ symbols.  For the purposes of this paper, we will not need to describe fusion associativity at this level.  However, if one wants to develop a full generalization of the algebraic theory of anyons for fracton phases -- which is \emph{not} the aim of this paper -- then presumably it would be necessary to introduce an $F$ symbol.

\subsection{Fusion theory with action of translation symmetry}
\label{subsec:fusion-translation}

The fusion theory we developed above is simply an Abelian group, and our considerations thus far are essentially the same as for Abelian point-like excitations in conventional iTO phases.  We could stop here, but the resulting fusion theory would not be very useful for fracton phases, because it contains no information about mobility of excitations.  Moreover, as we will see in explicit examples, ${\cal S}$ is not finitely generated for gapped fracton phases, and is thus a rather large mathematical object without much structure.

Therefore, we assume that our system is invariant under three-dimensional discrete translation symmetry; that is, it is invariant under the translation group $T \simeq \z^3$.  The translation group acts on ${\cal S}$; given a sector $s$, we can ask what sector results upon taking a representative excitation in $s$ and translating it by some amount.  Formally, for $t_{\ba} \in T$ and $s \in {\cal S}$, there is a function mapping $T \times {\cal S} \to {\cal S}$ that we write as $t_{\ba} s$. Here $\ba = (a_x, a_y, a_z)$ is a vector of integers labeling the translation, and we use multiplicative notation for $T$, so that $t_{\ba_1} t_{\ba_2} = t_{\ba_1 + \ba_2}$.  For $s, s_1, s_2 \in {\cal S}$, we assume on physical grounds that the following properties hold:
\begin{eqnarray}
t_{\ba_1} (t_{\ba_2} s) &=& t_{\ba_1 + \ba_2} s \\
t_{0} s &=& s \\
t_{\ba} (s_1 + s_2) &=& t_{\ba} s_1 + t_{\ba} s_2 
\end{eqnarray}
The first property says that translating in steps is the same as translating all at once.  The second property is obvious -- the trivial operation of no translation at all should leave the superselection sector unchanged.  The first two operations make the action of $T$ on ${\cal S}$ into a group action.  The third property says that the operations of fusion and translation commute -- that is, we can either first fuse two particles then translate the composite, or we can first translate each particle and then fuse them.  This property should hold because the corresponding representative excitations are the same in the two cases.  It follows from these properties that $t_{\ba} s = 0$ if and only if $s = 0$.

In fact, we can promote the translation group $T$ to its group ring with $\z$ coefficients, denoted $\zt$.  Elements of $\zt$ are formal integer linear combinations of elements of translations $t_{\ba}$, with multiplication acting in the obvious way,
\begin{equation}
(\sum_i n_i t_{\ba_i} ) ( \sum_j m_j t_{\ba_j} ) = \sum_{i, j} n_i m_j t_{\ba_i + \ba_j} \text{.}
\end{equation}
This is useful because there is a natural action of $\zt$ on ${\cal S}$.  For instance,
\begin{equation}
(t_{\ba_1} + t_{\ba_2}) s \equiv t_{\ba_1} s + t_{\ba_2} s \text{,}
\end{equation}
and
\begin{equation}
(t_{\ba_1} - t_{\ba_2}) s \equiv t_{\ba_1} s - t_{\ba_2} s \text{,}
\end{equation}
and finally
\begin{equation}
( n t_{\ba}) s \equiv \underbrace{t_{\ba} s + \cdots + t_{\ba} s}_{n \text{ times}} \text{.}
\end{equation}
This action makes ${\cal S}$ into a $\zt$-module.  We will sometimes refer to ${\cal S}$ and related objects as a module, and sometimes as a group.

We note that, beginning with the seminal work of Haah, $\zt$-modules have been used to analyze stabilizer code Hamiltonians realizing fracton phases \cite{haah11local,haah13commuting,yoshida,vijay16fracton, haah16algebraic}.  There, the structure appears upon representing Pauli operators or configurations of excitations in a stabilizer code as Laurent polynomials of  three variables $x$, $y$ and $z$, and noting that translation is the same as multiplication by a monomial.   Indeed, Ref.~\onlinecite{haah13commuting} introduced the same module ${\cal S}$ that we study here, referring to it as the set of topological charges, and used this to show that the energy cost of separating point-like excitations grows at most logarithmically with distance.  The emphasis there is different from the present work, focusing on general properties of stabilizer code models, rather than on characterizing quantum phases of matter.  Closer in perspective is Ref.~\onlinecite{haah16algebraic}, which contains a proof that ${\cal S}$ is independent of Hamiltonian representative for stabilizer codes (proposition V.12), and advocates that ${\cal S}$ be used as a characterization of stabilizer codes (section V.E).

\subsection{Fractons and sub-dimensional particles}
\label{subsec:mobility}

The fusion theory as developed allows us to study the mobility of excitations.  At the simplest level, it allows us to identify each superselection sector as a fracton, lineon, planon or fully mobile particle.  First, consider a sector $s \in {\cal S}$ and suppose that $t_{\ba} s = s$ for some $\ba \neq 0$.  Since in this case $s$ and $t_{\ba} s$ are in the same sector, there is some operator of bounded support that, acting on an excitation in sector $s$, has the same effect as translation by $\ba$.  We can think of this operator as a string operator, even though it does not necessarily have to be ``shaped like'' a string, because it has the effect of destroying an excitation in sector $s$ at some position $\br$, and re-creating an excitation in the same sector at $\br + \ba$.  Apart from destroying and creating excitations at these two ``endpoints'' of the string, no other excitations are created.

Because translation is a symmetry, an excitation and its image under $t_{\ba}$ have the same energy.  Therefore, if $t_{\ba} s = s$, excitations in the $s$ sector are mobile in the $\ba$ direction.  That is, they can hop by a displacement $\ba$ without changing energy, and this hopping is effected by the string operator discussed above.  

Excitations represented by each sector $s$ are mobile in zero, one, two or three dimensions, and the excitations are referred to as fractons, lineons, planons, and fully mobile particles, respectively.  More formally, these possibilities can be differentiated by considering the subgroup $T_s \subset T$ consisting of translations $t_{\ba}$ that take $s$ to itself; that is, translations satisfying $t_{\ba} s = s$.  Fractons have trivial $T_s$; that is, they have $t_{\ba} s \neq s$ for any $\ba \neq 0$.  For lineons, $T_s \simeq \z$, corresponding to translations along one direction.  Planons have $T_s \simeq \z^2$, corresponding to translations in a plane.  Finally, fully mobile particles have $T_s \simeq \z^3$.

\section{Fusion theories of some solvable fracton models}
\label{sec:examples}
\begin{figure}
    \centering
    \begin{subfigure}{0.9\columnwidth}
        \includegraphics[width=\textwidth]{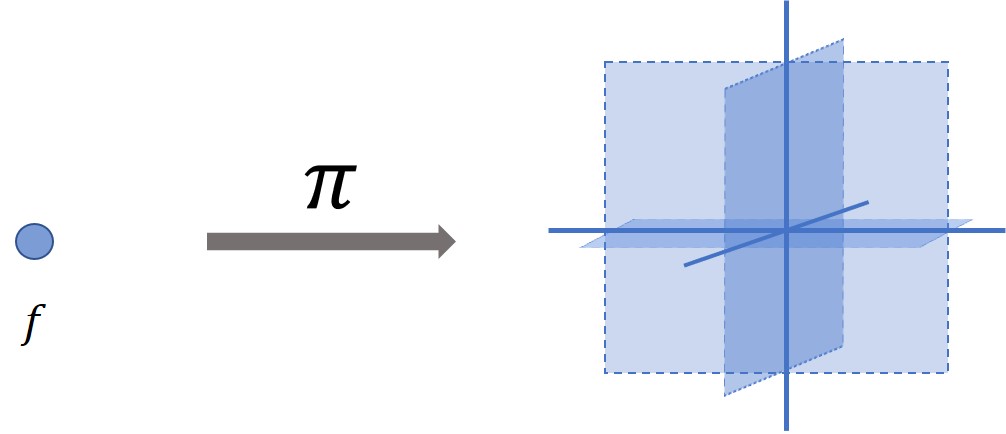}
    \end{subfigure}
        \vspace{0.25cm}
    \\ \hrule 
    \vspace{0.25cm}
    \begin{subfigure}{0.9\columnwidth}
        \includegraphics[width=\textwidth]{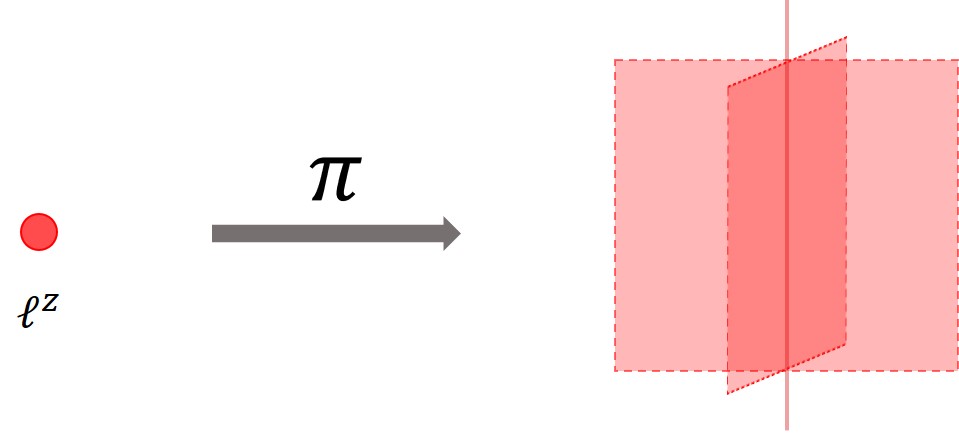}
    \end{subfigure}
    \caption{Graphical representation of the mapping between excitations in the X-cube model (left) and configurations of $\zz$ plane charges in the fusion theory (right).  In the top panel, a single cube excitation $f$ is a fracton that maps to a configuration of charges on three intersecting planes.  In the bottom panel, $\ell^z$ is a vertex excitation, and carries the charge on two intersecting planes.  It is apparent from the graphical representations that $\ell^z$ is a lineon able to move along the vertical axis, while $f$ is immobile.  We note that there are two distinct, independent types of plane charges, with cube excitations carrying one type, and vertex excitations carrying the other type (indicated by color online). The map $\pi$ is a map of $\zt$-modules defined in the text.}\label{fig:pimap}
\end{figure}

In this section, we describe the fusion theories of some exactly solvable gapped fracton models.  We first discuss the X-cube model, showing that its fusion theory decomposes into two sectors, and treating these in turn (Sec.~\ref{sec:xcube-fusion}).  We show that each sector is isomorphic to a certain submodule of a $\zt$-module of $\zz$ ``plane charges,'' in which a $\zz$ charge is attached to each lattice plane normal to the $x$, $y$ and $z$-directions.  The single cube excitations are fractons carrying the charge of three intersecting planes, while the vertex excitations are lineons with charges tied to two intersecting planes, as illustrated in Fig.~\ref{fig:pimap}.  

In Sec.~\ref{sec:xcube-pstring} we obtain the fusion theory of the X-cube model from a different point of view, starting from a system of decoupled $d=2$ toric code layers, and extending the corresponding fusion theory to implement the $p$-string condensation of Refs.~\onlinecite{ma17coupled,vijay17coupled}.  This treatment allows us to show that the X-cube and semionic X-cube models have the same fusion theory.  Finally, in Sec.~\ref{sec:checkerboard-fusion} we discuss the fusion theory for the checkerboard model.

\subsection{X-cube model}
\label{sec:xcube-fusion}

\begin{figure}[t]
  \includegraphics[scale=0.8]{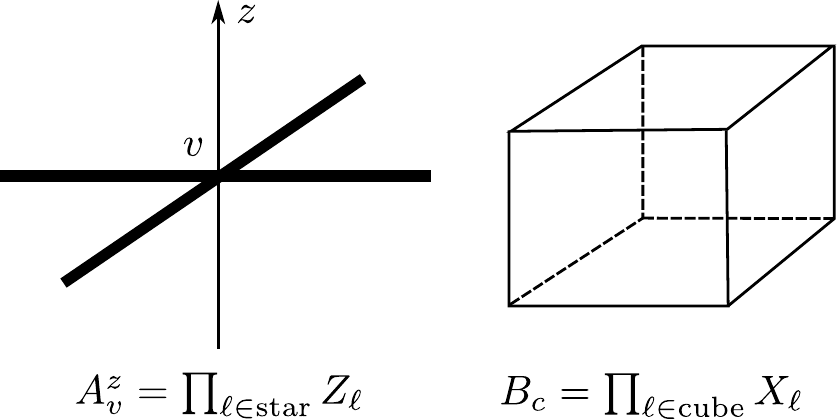}
 \caption{Operators in the X-cube model. (Left) $A_{v}^{z}$ is a product of $Z_{\ell}$ over the star of four thickened links. (Right) $B_{c}$ is the product of $X_{\ell}$ over the edges of an elementary cube.}
 \label{fig:Xcube}
 \end{figure}

We first consider the X-cube model \cite{vijay16fracton}, where one qubit is placed on each link $\ell$ of the $d=3$ simple cubic lattice.  We denote Pauli operators acting on each qubit by $X_{\ell}$ and $Z_{\ell}$. The Hamiltonian is
\begin{equation}
H_{\text{X-cube}} = -\sum_{v,\mu}A^{\mu}_{v} - \sum_{c}B_{c},
\end{equation}
where the first sum is over all vertices $v$ and directions $\mu=x,y,z$, and the second sum is over all elementary cubes $c$. As illustrated in Fig.~\ref{fig:Xcube}, $A^{\mu}_{v}$ is a product of $Z_{\ell}$ over the ``star'' of four links touching $v$ and lying in the plane normal to $\mu$, while $B_{c}$ is a product of $X_{\ell}$ over the $12$ links in the boundary of the cube $c$. All the terms in the Hamiltonian commute with one another, so the model is exactly solvable, and its energy eigenstates can be labeled by $a^{\mu}_v, b_c \in \{ \pm 1 \}$, the eigenvalues of $A^{\mu}_v$ and $B_c$, respectively.  Cube excitations with $B_c = -1$ are fractons, while vertex excitations with $A^{\mu}_v = -1$ are lineons.  

With periodic boundary conditions (\emph{i.e.} on a spatial 3-torus), the ground state and excited states are degenerate, with multiple states corresponding to a given choice of $a^{\mu}_v$ and $b_c$.  The degeneracy of the ground state grows sub-extensively with system size, that is the degeneracy scales like $\exp( c L)$, where $L$ is the linear system size.  While this degeneracy will not play an important role in our discussion, we note that the system has topological order, in the sense that the degenerate ground states cannot be distinguished by measurement of any local observables \cite{vijay16fracton}.
  
We are interested in configurations of excitations as specified by the sets of eigenvalues $\{ a^{\mu}_v \}$ and $\{ b_c \}$, in the limit of an infinite system, and with only finitely many $-1$ eigenvalues.  There is a local constraint $A^x_v A^y_v A^z_v = 1$ on the vertex terms, which implies $a^{x}_v a^y_v a^z_v = 1$.  Apart from this constraint, any configuration with finitely many $-1$ eigenvalues is realized by some eigenstate.\footnote{To see this, suppose that all the $-1$ eigenvalues lie within a ball of radius $R$ centered at the origin.  We can create such a state from a ground state by acting with string and membrane operators, where each string or membrane creates a single cube or vertex excitation within the ball, with the other excitations it creates lying outside the ball.  The infinite system limit corresponds to taking a limit of ``large'' string and membrane operators, so that all $-1$ eigenvalues outside the ball are pushed to infinity.}  Configurations of excitations form an Abelian group that we denote ${\cal E}$ and call the excitation group, where the group operation is multiplication of eigenvalues.  We pass to additive notation for ${\cal E}$ by writing $a^{\mu}_v = (-1)^{\alpha^{\mu}_v}$ and $b_c = (-1)^{\beta_c}$, with $\alpha^{\mu}_v, \beta_c \in \zz = \{0,1\}$.  Then elements of ${\cal E}$ are bit strings of the $\alpha^{\mu}_v$'s and $\beta_c$'s, with the group operation being element-by-element addition of the bit strings modulo two.

The excitation group decomposes as a direct sum ${\cal E} = {\cal E}_a \oplus {\cal E}_b$, where ${\cal E}_a = \bigoplus_v ( \zz \oplus \zz)$ consists of vertex-term excitations, and ${\cal E}_b = \bigoplus_c \zz$ consists of cube excitations.  (We note that the group of excitations at each vertex is $\zz \oplus \zz$ as a direct consequence of the constraint $a^{x}_v a^y_v a^z_v = 1$). Any locally createable energy eigenstate can be created from the ground state by acting with a finite product of Pauli operators, and since  $X_\ell$ operators only create vertex-term excitations, while $Z_\ell$ operators only create cube excitations, we also have the decomposition ${\cal L} = {\cal L}_a \oplus {\cal L}_b$, where ${\cal L}_{a} \subset {\cal E}_a$ consists of all excitations created by finite products of $X_\ell$ operators, and similarly for ${\cal L}_{b} \subset {\cal E}_b$.  

It follows that the fusion theory is also a direct sum; that is ${\cal S} = {\cal S}_a \oplus {\cal S}_b$, with ${\cal S}_a = {\cal E}_a / {\cal L}_a$, and similarly for ${\cal S}_b$.  Translation symmetry acts on ${\cal E}$, ${\cal L}$, and their direct summands ${\cal E}_a$, ${\cal E}_b$, ${\cal L}_a$ and ${\cal L}_b$, making these objects into $\zt$-modules, so that ${\cal S}$, ${\cal S}_a$ and ${\cal S}_b$ are also all $\zt$-modules.

We have thus shown that the fusion theory decomposes into a fusion theory associated with vertex excitations (${\cal S}_a$), and one associated with cube excitations (${\cal S}_b$).  This conclusion does not rely on any special properties of the X-cube model, but rather is a general property of commuting Pauli Hamiltonians where some terms are products only of $X_\ell$ and others are products only of $Z_\ell$.  Such decompositions are familiar from discrete Abelian gauge theories, where the fusion theory similarly decomposes into electric and magnetic sectors.

We refer to ${\cal S}_a$ and ${\cal S}_b$ as lineon and fracton fusion theories for the X-cube model, because their excitations can be obtained by fusing together lineons and fractons, respectively.  We treat the fracton and lineon fusion theories separately in the following two subsections.

\subsubsection{Fracton fusion theory of the X-cube model}
\label{subsubsec:Xcube-fracton}

Here we describe the fracton fusion theory ${\cal S}_b$ of the X-cube model, which is the theory of superselection sectors for the cube excitations.  We label cubes with integer coordinates $\br = (x,y,z)$.  The excitation group ${\cal E}_b$ is generated by elements $f(\br)$ satisfying $2 f(\br) = 0$, and a general element $e \in {\cal E}_b$ is written
\begin{equation}
e = \sum_{\br} \beta_{\br} f(\br) \text{,}
\end{equation}
where $\beta_{\br} \in \{0,1\}$ is the same as $\beta_c$ as defined above.  Translation symmetry acts in the obvious way,
\begin{equation}
t_{\ba} f(\br) = f(\br + \ba) \text{.}
\end{equation}

We specify the group ${\cal L}_b \subset {\cal E}_b$ in terms of its generators, which are simply the configurations of cube excitations created by acting with a single $Z_\ell$ operator on the ground state.  These are:
\begin{eqnarray}
&& f(\br) + f(\br + \hx) + f(\br + \hy) + f(\br + \hx + \hy)\nonumber  \\
&&  f(\br) + f(\br + \hx) + f(\br + \hz) + f(\br + \hx + \hz)  \nonumber \\
&&  f(\br) + f(\br + \hy) + f(\br + \hz) + f(\br + \hy + \hz) \nonumber \text{,}
\end{eqnarray}
where $\hx, \hy$ and $\hz$ are the usual Cartesian unit vectors.

Our discussion so far is enough to define ${\cal S}_b = {\cal E}_b / {\cal L}_b$, but we will go further and obtain a simple description of ${\cal S}_b$ that will allow us to understand the fusion and mobility properties of all excitations that can be obtained from single cube excitations.  In order to do this, we recall that the number of cube excitations in every $\{ 100 \}$ lattice plane is conserved modulo two \cite{vijay16fracton}.\footnote{We recall that the Miller index notation $\{ k l m \}$ denotes a family of symmetry-equivalent lattice planes, while $(k l m)$ denotes a family of lattice planes all with the same orientation.  So, for instance, in a cubic crystal $(100)$ and $(010)$ are distinct sets of lattice planes both belonging to the family of $\{100 \}$ planes.} That is, there is no local process that adds or removes a single cube excitation to a plane, as is easily seen from the form of the generators of ${\cal L}_b$.  

This motivates us to define a group of plane charges ${\cal P} = \bigoplus_p \zz$, where the direct sum is over $\{ 100 \}$ planes.  The generator for the $\zz$ summand associated with the $xy$ plane with $z$-coordinate $z$ is written $P_{xy}(z)$, and similarly we introduce generators $P_{xz}(y)$ and $P_{yz}(x)$ for the other $\zz$ summands. Translation symmetry acts on ${\cal P}$ as a translation of the lattice planes.  That is, for a translation $t_{\ba} = (a_x, a_y, a_z)$, we have $t_{\ba} P_{xy}(z) = P_{xy}(z + a_z)$, and similarly for the other generators.  This makes ${\cal P}$ into a $\zt$-module.

We will see that ${\cal S}_b$ is isomorphic to a certain submodule of ${\cal P}$.  
The plane charge of any excitation ${\cal E}_b$ is computed by a map $\pi : {\cal E}_b \to {\cal P}$, defined by its action on generators:
\begin{equation}
f(x,y,z) \mapsto  P_{yz}(x)  + P_{xz}(y) + P_{xy}(z) \text{.}
\label{eqn:fractonmap} 
\end{equation}
This expresses the fact that a single cube excitation carries plane charges in three perpendicular lattice planes; the top panel of Fig.~\ref{fig:pimap} is a graphical illustration.
This map commutes with the action of translation and so is a map between $\zt$-modules.  The observation that locally createable excitations have trivial plane charges is then written $\pi(e) = 0$ for any $e \in {\cal L}_b$.  That is, ${\cal L}_b \subset \operatorname{ker} \pi$.  Therefore, remembering that ${\cal S}_b = {\cal E}_b / {\cal L}_b$, $\pi$ induces a map $\pi_S : {\cal S}_b \to {\cal P}$.  In fact, as shown in Appendix~\ref{app:Xcube-fracton}, $\operatorname{ker} \pi = {\cal L}_b$, which implies that $\pi_S$ is injective.  Physically this means that different superselection sectors can be uniquely labeled by configurations of plane charges.

Because $\pi_S$ is injective, we have
\begin{equation}
{\cal S}_b \simeq \pi_S ( {\cal S}_b ) \equiv {\cal P}_b \subset {\cal P} \text{.}
\end{equation}
This defines a submodule ${\cal P}_b \subset {\cal P}$ that is isomorphic to ${\cal S}_b$, and from now on we identify ${\cal S}_b$ with ${\cal P}_b$.  This gives a description of the superselection sectors in terms of plane charges.  ${\cal P}_b$ is generated by
$\pi[ f(\br)] = P_{yz}(x)  +P_{xz}(y) + P_{xy}(z)$.   It is important to note that ${\cal P}_b$ is smaller than ${\cal P}$; that is, some configurations of plane charges are not allowed.  In fact, if ${\cal P}_b$ were the same as ${\cal P}$, then every cube excitation would necessarily be a composite of planons, which is not the case.  It is thus essential to characterize the subgroup ${\cal P}_b \subset {\cal P}$, which we now do.

A general element $ p \in {\cal P} $ can be written
\begin{eqnarray}
 p &=& \sum_{x} Q_{yz}(x) P_{yz}(x) + \sum_{y} Q_{xz}(y) P_{xz}(y)  \nonumber \\ &+& \sum_{z} Q_{xy}(z) P_{xy}(z) \text{,}  \label{eqn:general-p}
 \end{eqnarray}
 where $Q_{yz}(x) \in \{0,1\}$, and similarly for the other coefficients.  We define a triple of integers characterizing $p$:
 \begin{eqnarray}
 k_p &=& \sum_x Q_{yz}(x) \\
 l_p &=& \sum_y Q_{xz}(y) \\
 m_p &=& \sum_z Q_{xy}(z) \text{.}
 \end{eqnarray}
 In Appendix~\ref{app:Xcube-fracton}, it is shown that  that $p \in {\cal P}_b$ if and only if  $k_p$, $l_p$ and $m_p$ are all even or all odd.  This gives an explicit and simple description of ${\cal S}_b$ in terms of plane charges.

Now we derive some consequences of this result.  First, we see that the cube excitations are fractons by the definition given in Sec.~\ref{subsec:mobility}.  That is,
\begin{equation}
 \pi [ t_{\ba} f(\br) ] \neq \pi [ f(\br) ] \text{,} \label{eqn:xcube-immobile}
 \end{equation}
for $\ba \neq 0$, which is easily seen by explicitly evaluating the maps on the left- and right-hand sides.  Equivalently, again for $\ba \neq 0$, we have
\begin{equation}
\pi [ t_{\ba} f(x,y,z) + f(x,y,z)]  \neq 0 \text{,}
\end{equation}
which is the statement that a pair of cube excitations is never locally createable, \emph{i.e.} there are no string operators that create a pair of cube excitations.  We note that it has already been established in Ref.~\onlinecite{vijay16fracton} that a pair of cube excitations cannot be created by any string operator.  By obtaining this result from the fusion theory, our treatment makes it clear that this is a universal property of a quantum phase of matter, at least when translation symmetry is present.

We can also use the fusion theory to discuss composites of cube excitations.  For instance, it is known that certain composites of two cube excitations are planons.  For instance, the excitation $f(\br) + f(\br + n \hz)$ can move in the $xy$-plane, as we can see from
\begin{equation}
\pi [ f(\br) + f(\br + n \hz) ] = P_{xy}(z) + P_{xy}(z+n) \text{,}
\end{equation}
where $\br = (x,y,z)$; the right-hand side is clearly invariant under any translation in the $xy$-plane.

More generally, we can use the fracton fusion theory to study arbitrary composites of cube excitations.  We refer to elements of ${\cal S}_b$ with $k_p, l_p, m_p$ all odd (all even) as odd (even) elements.  The sum of two even elements  or two odd elements is even, and summing an even element with an odd element gives an odd element.  Therefore, there is a homomorphism from ${\cal S}_b$ to $\zz$ that maps odd elements to $1 \in \zz$ and even elements to $0 \in \zz$.  It is easy to see that all even excitations can be obtained by fusing together a finite number of planons of the form ${f(\br) + f(\br + \hx)}$, ${f(\br) + f(\br + \hy)}$ and ${f(\br) + f(\br + \hz)}$.  Moreover, an excitation is a planon if and only if exactly two of $k_p, l_p, m_p$ are zero, and is a lineon if and only if exactly one of $k_p, l_p, m_p$ is zero.  (We note that these lineons are not intrinsic lineons, \emph{i.e.} they are composites of planons.)  Therefore, only even excitations can be lineons or planons, while all odd excitations are fractons.  While there are even fractons, these can all be obtained as bound states of planons.  Odd excitations, on the other hand, are \emph{intrinsic} fractons, by which we mean they cannot be obtained as composites of excitations with higher mobility \cite{song18twisted,prem18cagenet}.

%Therefore, only \textcolor{Green}{even excitations can be planons or their bound states (``even'' fractons, lineons)}, while all odd excitations are fractons.  \textcolor{Green}{In particular, even fractons are all bound states of planons.}  Odd excitations, on the other hand, are \emph{intrinsic} fractons, by which we mean they cannot be obtained as composites of excitations with higher mobility \cite{song18twisted,prem18cagenet}.

Ref.~\onlinecite{shirley2019fractional} introduced the notion of quotient superselection sectors (QSS), which are obtained by viewing two sectors related by fusing planons as equivalent.  In the language of our fusion theory, the theory of QSS is a $\zt$-module $Q({\cal S}_b)$ obtained by constructing the submodule   $P({\cal S}_b) \subset {\cal S}_b$ generated by all planon excitations, and forming the quotient $Q({\cal S}_b) = {\cal S}_b / P({\cal S}_b)$.  In the present case, $P({\cal S}_b)$ is simply the subgroup of even superselection sectors, and $Q({\cal S}_b) \simeq \zz$, where the quotient map is the same homomorphism from ${\cal S}_b$ to $\zz$ discussed above.  This tells us that any two odd excitations are related by fusing planons, which is also easy to see directly.

\subsubsection{Lineon fusion theory of the X-cube model}
\label{subsubsec:Xcube-lineon}

We now turn to the fusion theory ${\cal S}_a$ in the X-cube model, which we refer to as the lineon fusion theory, because it describes the lineon vertex excitations and their composites.  The excitation group ${\cal E}_a$ is generated by elements $\ell^{\mu}(\br)$ satisfying $2 \ell^{\mu}(\br) = 0$ and 
\begin{equation}
 \ell^x(\br)   + \ell^y(\br) +\ell^z(\br)  = 0 \text{,} \label{eqn:cage-reln} 
 \end{equation}
 where $\br = (x,y,z)$ is the position vector of a vertex with $x,y,z \in \z$.
 A general excitation $e \in {\cal E}_a$ can be expressed
 \begin{equation}
 e = \sum_{\br, \mu} n_{\mu}(\br) \ell^{\mu}(\br) \text{,}  \label{eqn:general-lineon}
 \end{equation}
 where $n_{\mu}(\br) \in \{0 ,1\}$, and where for a fixed position $\br$, we can take at most one of the $n_{\mu}(\br)$ to be equal to unity.  The quantities $n_{\mu}(\br)$ are related to $\alpha^{\mu}_v$ introduced above by $n_{\mu}(\br) = \prod_{\nu \neq \mu} \alpha^{\nu}_{\br}$.

The submodule of locally createable excitations $ {\cal L}_a \subset {\cal E}_a $ is generated by elements of the form
  \begin{eqnarray}
&& \ell^{x}(\br) + \ell^{x}(\br + \hx) \text{,} \nonumber  \\
&& \ell^{y}(\br) + \ell^{y}(\br + \hy) \text{,} \nonumber \\
&& \ell^{z}(\br) + \ell^{z}(\br + \hz) \nonumber \text{,}
\end{eqnarray} 
where these are the configurations of excitations created by acting with $X_\ell$ on some link of the cubic lattice.  From the form of the generators, we can observe that, for instance, the total number of $\ell^x$ plus $\ell^y$ vertex excitations in any $xy$-plane is conserved modulo two, with analogous statements holding for $xz$ and $yz$ planes.  This motivates us to again introduce the group ${\cal P}$ of plane charges as above in Sec.~\ref{subsubsec:Xcube-fracton}.  Here, we define a map  $ \pi : {\cal E}_a \to {\cal P} $ by
\begin{eqnarray}
&& \ell^{x}(x,y,z) \mapsto P_{xz}(y) + P_{xy}(z)  \nonumber \\ 
&& \ell^{y}(x,y,z) \mapsto P_{yz}(x) + P_{xy}(z)   \label{eqn:Sa-generators}  \\ 
&& \ell^{z}(x,y,z) \mapsto P_{yz}(x) + P_{xz}(y)  \nonumber  \text{,}
\end{eqnarray} 
which is again a map between $\zt$-modules, illustrated in the bottom panel of Fig.~\ref{fig:pimap}.

In Appendix~\ref{app:Xcube-lineon} it is shown that ${\cal L}_a = \operatorname{ker} \pi$, so as before, $\pi$ induces an injective map $\pi_{S}: {\cal S}_a \rightarrow {\cal P}$, and we identify ${\cal S}_a \simeq \pi_S({\cal S}_a) \equiv {\cal P}_a \subset {\cal P}$.  We need to characterize the subgroup ${\cal P}_a \subset {\cal P}$.  ${\cal P}_a$ is generated by elements of the form given in Eq.~(\ref{eqn:Sa-generators}).  As before, any $p \in {\cal P}$ can be written in the form Eq.~(\ref{eqn:general-p}), and characterized by the integers $k_p, l_p$ and $m_p$.  In Appendix~\ref{app:Xcube-lineon} we show that $p \in {\cal P}_a$ if and only if $k_p + l_p + m_p = 0 \mod 2$, which gives the desired characterization.

From this description of the lineon fusion theory, we see that a single vertex excitation $\ell^{\mu}(\br)$ is indeed a lineon.  For instance, $\ell^x(x,y,z)$ maps into $P_{xz}(y) + P_{xy}(z) \in {\cal P}_a$, which is invariant under translations in the $x$-direction, but 
\begin{equation}
\pi [ \ell^x(x,y,z) ] \neq \pi [ t_{\ba}  \ell^x(x,y,z)  ] \text{,}
\end{equation}
for any $\ba = (a_x,a_y,a_z)$ with $a_y \neq 0$ or $a_z \neq 0$.
We can also see that certain composites of two vertex excitations are planons \cite{vijay16fracton}.  For instance, $\ell^x(x,y,z) + \ell^x(x,y,z+1)$ maps to $P_{xy}(z) + P_{xy}(z+1)$, which is clearly invariant under translations in the $xy$-plane, but not under translations in the $z$-direction.  

More generally, just as for the fracton fusion theory, excitations are planons when exactly two of $k_p, l_p$ and $m_p$ are zero, and they are lineons when exactly one of $k_p, l_p$ and $m_p$ is zero.   For planons, the non-zero integer among $\{ k_p, l_p, m_p \}$ is even. Intrinsic lineons, including single vertex excitations, are those for which the non-zero integers among $\{ k_p, l_p, m_p \}$ are odd, as these excitations cannot be obtained as composites of higher-mobility planons.  There is a homomorphism from ${\cal S}_a$ to $\zz \oplus \zz$ defined by $(k_p, l_p, m_p) \mapsto (k_p \mod 2, l_p \mod 2, m_p \mod 2)$, and the kernel of this homomorphism is precisely the $\zt$-module $P({\cal S}_a)$ generated by planons.  The theory of QSS is then given by $Q({\cal S}_a) = {\cal S}_a / P({\cal S}_a) \simeq \zz \oplus \zz$, where the three non-trivial elements correspond to fundamental lineons mobile in the $x$, $y$ and $z$ directions, the same result obtained in Ref.~\onlinecite{shirley2019fractional}.

\subsection{X-cube and semionic X-cube fusion theory from $p$-string condensation}
\label{sec:xcube-pstring}

It has been shown that the X-cube model can be obtained  from three perpendicular stacks of decoupled $d=2$ toric code models upon suitably coupling the layers \cite{ma17coupled,vijay17coupled}.  In this construction, a square lattice $d=2$ toric code model is placed on each $\{100\}$ square lattice layer of the simple cubic lattice, so that two qubits reside on each link.  The coupling between the layers creates $p$-strings, or particle strings, which are string-like objects built from the $m$ particle plaquette excitations of the toric code layers.  It was argued that condensing the $p$-strings results in the phase of the X-cube model.  We refer the reader to Refs.~\onlinecite{ma17coupled,vijay17coupled} for more details.

\begin{figure}
    \centering
    \begin{subfigure}{0.40\columnwidth}
        \includegraphics[width=\textwidth]{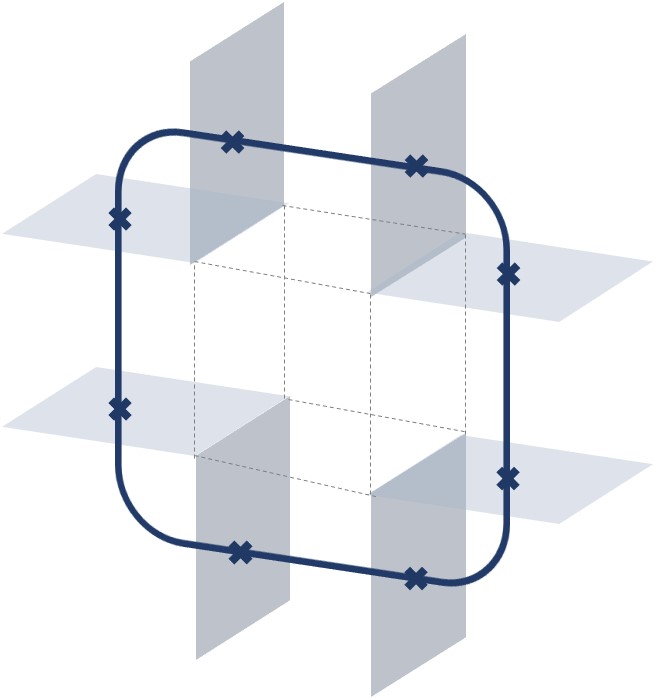}
        \caption{}
    \end{subfigure}
    \hfill \vrule \hfill
    \begin{subfigure}{0.25\columnwidth}
        \includegraphics[width=\textwidth]{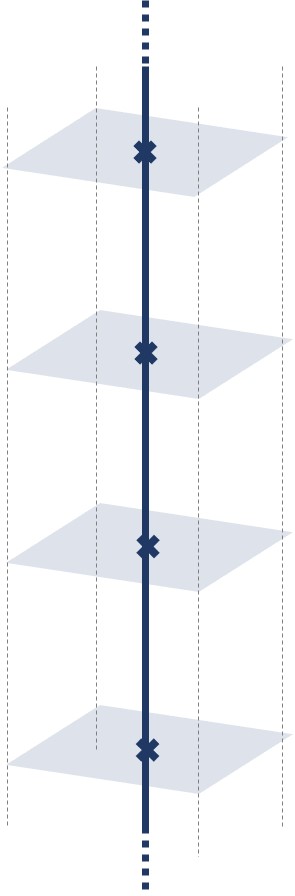}
        \caption{}
    \end{subfigure}
    \caption{Illustration of two closed $p$-string configurations of $m$ particle excitations in a model of decoupled $d=2$ toric code layers on the simple cubic lattice.  Panel (a) shows a finite closed $p$-string, while (b) is an infinite closed $p$-string, which corresponds to an element $f^-_z(\br) + f^+_z(\br) \in {\cal E}^{{\rm ext}}_m$.  The gray-shaded plaquettes are $m$-particle excitations, and a segment of blue $p$-string is shown intersecting transversely to each $m$-particle plaquette.  The $p$-strings here are infinite (no endpoints) because every cube has an even number of $m$-particle excitations on its faces.}\label{fig:pstrings}
\end{figure}

Here, we would like to describe $p$-strings and their condensation within the framework of our fusion theory, both for the X-cube model and the semionic X-cube model of Ref.~\onlinecite{ma17coupled}.  We focus with the X-cube model and discuss its semionic cousin at the end of the section.  Starting with decoupled toric code layers, we say that an excitation is a closed $p$-string if it consists only of $m$ particles (plaquette excitations), and if there are an even number of $m$ particle excitations on the six faces of each elementary cube of the lattice.  As shown in Fig.~\ref{fig:pstrings}, such configurations can be viewed as Ising string configurations on the dual cubic lattice, by drawing a perpendicular line through each plaquette occupied by an $m$ particle.

The fusion theory for the system of decoupled toric code layers is ${\cal S} = {\cal S}_e \oplus {\cal S}_m$, where ${\cal S}_e$ is the fusion theory of $e$ particle excitations (vertex excitations) and ${\cal S}_m$ that of $m$ particle excitations.  We have
\begin{eqnarray}
{\cal S}_e = \oplus_{{\rm layers}} \, \zz \\
{\cal S}_m = \oplus_{{\rm layers}} \, \zz \text{,}
\end{eqnarray}
where the direct sums are over all the decoupled toric code layers.  The $\zz$ generator corresponding to an $e$ particle ($m$ particle) in the layer normal to the $z$-axis with $z = z_0$ is written $e_z(z_0)$ [$m_z(z_0)$], and similarly for the other two orientations of layers.  Translation symmetry acts in the obvious way.

It is easy to see that finite closed $p$-strings are locally createable and thus are trivial excitations.  Therefore, in order to describe $p$-string condensation from the perspective of the fusion theory, we need to work with infinite $p$-strings (see Fig.~\ref{fig:pstrings}b).  But these objects are not contained within the fusion theory as it stands, because we only consider excitations of bounded support.  We thus need to extend the fusion theory to include certain excitations of unbounded support, and in order to do this we go back to the description of excitations in terms of the excitation group, or $\zt$-module, ${\cal E} = {\cal E}_e \oplus {\cal E}_m$.

\begin{figure}
  \includegraphics[width=0.6\columnwidth]{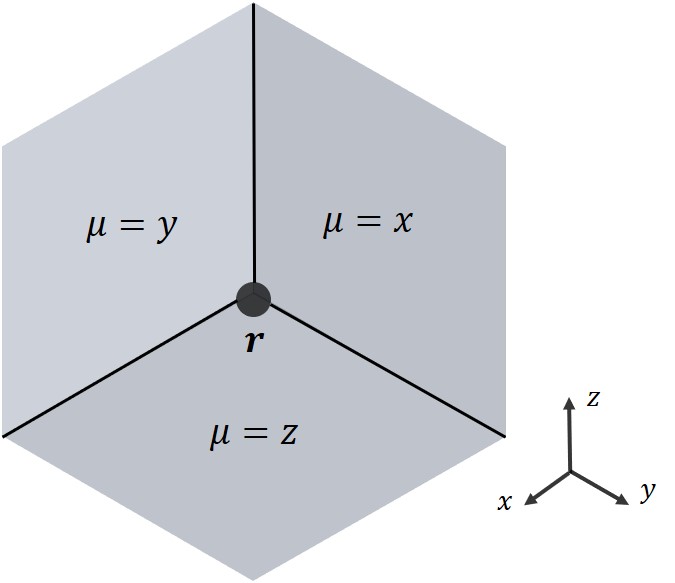}
 \caption{Illustration of the labeling of cubic lattice plaquettes used in the discussion of $p$-string condensation.  Each plaquette is labeled by a pair $(\br, \mu)$, where $\mu$ is the normal direction, and $\br$ is the vertex at the corner of the plaquette as shown.}
 \label{fig:pstring-labeling}
 \end{figure}

It is enough to focus on ${\cal E}_m$, because the $p$-strings are built from $m$ particles.  We have ${\cal E}_m = \oplus_p \zz$, where the direct sum is over all square plaquettes of the cubic lattice.  We label plaquettes $p$ by the position of one corner $\br = (x,y,z)$ and the direction of the normal vector $\mu = x,y,z$, and denote the corresponding generator by $m_{\mu}(\br)$ (see Fig.~\ref{fig:pstring-labeling}).  Pairs of $m$ particles on neighboring plaquettes lying within the same layer can be created locally, and ${\cal L}_m \subset {\cal E}_m$ is generated by $m_z(\br) + m_z(\br + \hx)$, $m_z(\br) + m_z(\br + \hy)$, and the corresponding elements for other orientations of plaquettes.

By definition, elements of ${\cal E}_m$ are excitations of bounded support.  We now define an extended excitation $\zt$-module ${\cal E}^{{\rm ext}}_m$ with ${\cal E}_m \subset {\cal E}^{{\rm ext}}_m$ that includes infinite $p$-strings, which are unbounded excitations.  ${\cal E}^{{\rm ext}}_m$ is generated by elements $f^{\pm}_\mu(\br)$, where $\mu = x,y,z$, which are open infinite $p$-strings.  For instance, we define
\begin{eqnarray}
f^+_z(x_0, y_0, z_0) &=& \sum_{z = z_0 + 1}^{\infty} m_z(x_0, y_0, z) \\
f^-_z(x_0, y_0, z_0)  &=& \sum_{z = -\infty}^{z_0} m_z(x_0, y_0, z)  \text{,}
\end{eqnarray}
with similar expressions for the other orientations.  The offset of $+1$ in the definition of $f^+_z(x_0, y_0, z_0)$ ensures that, upon $p$-string condensation, the corresponding fracton excitation has coordinates $(x_0, y_0, z_0)$.  The generators of ${\cal E}_m$ can be obtained from these generators of ${\cal E}^{{\rm ext}}_m$; for instance, 
\begin{equation}
m_z(\br) = f^-_z(\br) + f^-_z(\br - \hz) \text{.}
\end{equation}

${\cal E}^{{\rm ext}}_m$ contains both finite closed $p$-strings, and infinite closed $p$-strings such as $f^-_z(\br) + f^+_z(\br)$ and $f^-_z(\br) + f^+_y(\br)$.  It is easy to see that fusing two closed $p$-strings (finite or infinite) gives another closed $p$-string, so that closed $p$-strings form a submodule of ${\cal E}^{{\rm ext}}_m$.  

We now consider the extended fusion theory ${\cal S}^{{\rm ext}}_m = {\cal E}^{{\rm ext}}_m / {\cal L}_m$.  It is important to note that, while we modified the excitation group, we do not modify the group of locally createable excitations ${\cal L}_m$, because we still want to consider two excitations equivalent only when they are related by an operator of bounded support.  Consider the infinite open $p$-string $f^-_z(x_0,y_0,z_0)$, which comes with a semi-infinite tail extending in the $-z$-direction with transverse position $(x_0,y_0)$.  The direction of the tail and its transverse position at infinity cannot be changed by acting with an operator of bounded support, and this information is thus preserved upon taking the quotient by ${\cal L}_m$.  Moreover, the coordinate $z_0$ is also robust under acting with bounded operators, because changing $z_0$ corresponds to creating or destroying $m$ particles.  Therefore, all the $f^\pm_{\mu}(\br)$ generators remain distinct upon taking the quotient by ${\cal L}_m$.  That is 
\begin{equation}
q_l [ f^\sigma_{\mu}(\br) ] = q_l [ f^{\sigma'}_{\mu'}(\br') ]
\end{equation}
if and only if $\sigma = \sigma'$, $\mu = \mu'$ and $\br = \br'$, where $q_l$ is the quotient map and $\sigma, \sigma' \in \{+, - \}$.

The extended fusion theory ${\cal S}^{{\rm ext}}_m$ looks almost like ${\cal S}_b$, the fracton fusion theory of the X-cube model, with $q_l [ f^\pm_{\mu}(\br) ]$ playing a similar role to $\pi[ f(\br) ]$.  However, there is the important and obvious difference that, here, the generators depend on the direction of the tail, while there are no such labels of the generators in ${\cal S}_b$.  Indeed, in ${\cal S}^{{\rm ext}}_m$ the generators group into six classes if we treat the action of translation as an equivalence operation, while for ${\cal S}_b$ all the generators are related by translation.

Upon $p$-string condensation, closed $p$-strings can disappear into and appear from the vacuum, and should be identified with trivial excitations.  Mathematically, we describe this via the quotient ${\cal S}^{{\rm ext}}_m / \Sigma$, where $\Sigma \subset {\cal S}^{{\rm ext}}_m$ is the submodule of closed $p$-strings.  Denoting this quotient map by $q_\sigma$,  we identify ${\cal S}^{{\rm ext}}_m / \Sigma$ with the fracton fusion theory ${\cal S}_b$ via 
\begin{equation}
q_\sigma \circ q_l [ f^{\pm}_{\mu}(\br) ] = \pi[ f(\br) ] \text{.}
\end{equation}

It still remains to describe the lineon fusion theory ${\cal S}_a$ from this perspective.  Here, following Refs.~\onlinecite{ma17coupled,vijay17coupled}, we need to consider statistics of excitations in ${\cal S}_e$ with the $p$-string condensate.  There is a statistical phase of $-1$ when a single $e$-particle excitation is braided around a $p$-string, so these excitations are confined.  However, bound states of pairs of $e$ particles in perpendicular planes have trivial statistics with $p$-strings, and these excitations correspond to the elementary lineons (vertex excitations) of ${\cal S}_a$.

This approach can also be used to obtain the fusion theory of the semionic X-cube model introduced in Ref.~\onlinecite{ma17coupled}.  That model is constructed via $p$-string condensation starting from decoupled layers of $d=2$ doubled semion string-net models \cite{levin05stringnet}.  The non-trivial excitations of each doubled semion layer are a bosonic flux $m$, and semions $s$ and $s' = s + m$.  The single-layer fusion theory is $\zz \oplus \zz$.  The $p$-strings are built from the $m$-particles, and we can obtain the fusion theory of the semionic X-cube model exactly as above.  While the semionic X-cube model is defined on a decorated version of the cubic lattice, this does not play an important role (as explained below), and the fusion theory is identical to that of the X-cube model, with fractons arising at ends of open $p$-strings, and lineons arising as bound states of semions in two perpendicular layers.

In more detail, in the construction of Ref.~\onlinecite{ma17coupled}, each doubled semion layer is a square-octagon lattice, and these layers are assembled into a modified simple cubic lattice; each cube is truncated at its corners, resulting in a lattice composed of truncated cubes with octagonal and triangular faces, and octahedra centered on cubic lattice vertices.  Every $m$-particle excitation can be moved to an octagonal plaquette, and closed $p$-strings are defined to be excitations composed only of $m$-particles, where $m$ particles occupy an even number of the octagonal faces of each truncated cube.  Apart from these minor modifications, the analysis proceeds exactly as in the case of the X-cube model.

It has been shown that the X-cube and semionic X-cube models realize the same foliated fracton phase \cite{shirley2019fractional}.  However, despite having the same fusion theory, in Sec.~\ref{sec:2Xcube} we prove these models are in distinct translation-invariant fracton phases, because their statistical properties are different.

\subsection{Fusion of fractons in the checkerboard model}
\label{sec:checkerboard-fusion}

\begin{figure}
  \includegraphics[width=\columnwidth]{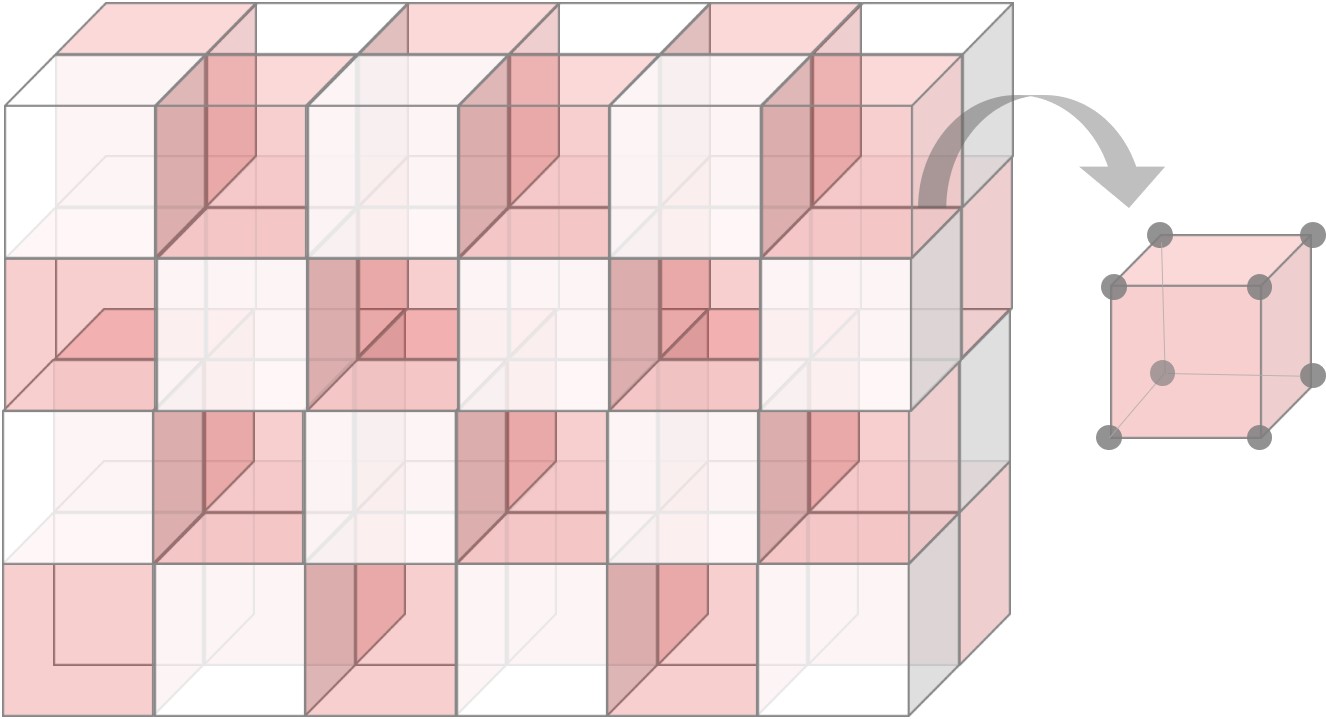}
 \caption{The shaded cubes form the A sublattice in the checkerboard model.  The $A_c$ and $B_c$ terms in the Hamiltonian are products of $Z_v$ and $X_v$, respectively, over the eight vertices at the corners of each A-sublattice cube.}
 \label{fig:CBlattice}
 \end{figure}

We now consider the fracton or cube excitations of the checkerboard model \cite{vijay16fracton}. In the checkerboard model, qubits are placed on the sites $v$ of a $d=3$ cubic lattice, with Pauli operators denoted by $X_{v}$ and $Z_{v}$.  We divide the elementary cubes into A and B sublattices, so that each sublattice consists of edge-sharing cubes forming a ``checkerboard'' pattern (see Fig.~\ref{fig:CBlattice}).  Giving the cube centers integer coordinates $\br = (x,y,z)$, the A sublattice is defined by requiring $x+y+z = 0 \mod 2$.  We denote the set of points in the A sublattice by $\Lambda$, and $\Lambda$ is a face centered cubic lattice with primitive translation vectors $\ba_1 = (0,1,1)$, $\ba_2 = (1,0,1)$ and $\ba_3 = (1,1,0)$. The Hamiltonian is invariant under this translation symmetry and is
\begin{equation}
H_{\text{checkerboard}} = -\sum_{c \in \Lambda} A_c -\sum_{c \in \Lambda} B_c \text{,}
\label{CBHamiltonian}
\end{equation}
where $A_c = \prod_{v \in c} Z_v$ and $B_c = \prod_{v \in c} X_v$.  All the terms in the Hamiltonian commute with one another, so the model is exactly solvable, with energy eigenstates labeled by eigenvalues $a_c$ and $b_c$ of $A_c$ and $B_c$, respectively.
In a ground state, $a_c = b_c = 1$ for all $c \in \Lambda$.  Single cube excitations, where one of $a_c$ or $b_c$ is equal to $-1$, are fractons. 

As in the X-cube model, the fusion theory decomposes as ${\cal S} = {\cal S}_a \oplus {\cal S}_b$. The model has an obvious self duality where $X_v$ and $Z_v$ are exchanged, so ${\cal S}_a$ and ${\cal S}_b$ are isomorphic as $\zt$-modules, and it is enough to focus on $a_c = -1$ excitations and ${\cal S}_a$.

The excitation group is ${\cal E}_a = \bigoplus_{\br \in \Lambda} \zz$, with the $\zz$ summand for the cube at $\br$ denoted by $f(\br)$.  The generators of ${\cal L}_a \subset {\cal E}_a$ are elements of the form
\begin{eqnarray}
&& f(\br) + f(\br + \hx + \hy) + f(\br + \hy + \hz)  + f(\br + \hx + \hz) \text{,} \nonumber  \\
&& f(\br) + f(\br + \hx + \hy) + f(\br + \hy - \hz)  + f(\br + \hx - \hz) \nonumber \text{.}
\end{eqnarray}
Again we introduce the group ${\cal P}$ of $\zz$ plane charges in $\{ 100 \}$ planes, and, exactly as in the fracton fusion theory of the X-cube model, Eq.~(\ref{eqn:fractonmap}) defines a map $\pi : {\cal E}_a \to {\cal P}$.

It is true that $\operatorname{ker} \pi = {\cal L}_a$; it is obvious that ${\cal L}_a \subset \operatorname{ker} \pi$, and the reverse inclusion is shown in Appendix~\ref{app:checkerboard}.  Therefore $\pi$ induces an injective map $\pi_S : {\cal S}_a \to {\cal P}$, and as before we identify ${\cal S}_a$ with $\pi_S({\cal S}_a) = {\cal P}_a \subset {\cal P}$.

Our next task is to characterize the submodule ${\cal P}_a \subset {\cal P}$.  A general element $p \in {\cal P}$ can be expressed as in Eq.~(\ref{eqn:general-p}).  First, we define integers characterizing $p$ by
\begin{eqnarray}
k^o_p &=& \sum_{x \,\, \text{odd}} Q_{yz}(x)  \\
k^e_p &=& \sum_{x \,\, \text{even}} Q_{yz}(x)  \text{,}
\end{eqnarray}
and similarly for $l^{o,e}_p$ (with sums over odd and even $y$), and $m^{o,e}_p$ (with sums over odd and even $z$).
We define a submodule $\tilde{\cal P} \subset {\cal P}$ by imposing the constraints
\begin{eqnarray}
k^o_p + l^o_p + m^o_p &=& 0 \mod 2 \label{eqn:check1} \\
k^o_p + l^e_p + m^e_p &=& 0 \mod 2 \\
k^e_p + l^o_p + m^e_p &=& 0 \mod 2 \\
k^e_p + l^e_p + m^o_p &=& 0 \mod 2  \label{eqn:check4} \text{.}
\end{eqnarray}
In Appendix~\ref{app:checkerboard} it is shown that $\tilde{\cal P} = {\cal P}_a$, so these constraints provide the desired characterization of ${\cal P}_a$.

In Ref.~\onlinecite{shirley18checkerboard}, it was shown that the checkerboard model is equivalent to two copies of the X-cube model.  More precisely, an explicit local unitary was found that maps the checkerboard model to two copies of the X-cube model plus trivial ancilla qubits.  This mapping does not respect the full translation symmetry of the checkerboard model, but instead breaks it down to the lower symmetry of translations by $\ba = (a_x,a_y,a_z)$ with even $a_x,a_y,a_z$.  The equivalence between the checkerboard model and two copies of the X-cube model should thus be manifest in the fusion theory if we also break the translation symmetry accordingly, and we now show that this is indeed the case.

We will show that under the reduced translation symmetry, ${\cal S}_a = {\cal S}_f \oplus {\cal S}_\ell$ as a direct sum of $\zt$-modules, where ${\cal S}_f$ and ${\cal S}_\ell$ are isomorphic to the fracton and lineon fusion theories of the X-cube model, respectively.  We let $\bR = (x,y,z)$ with $x,y,z$ even, so that $\bR$ labels unit cells under the reduced translation symmetry.  There are four cube centers in each unit cell, and we choose a set of generators for ${\cal E}_a$ as follows:
\begin{eqnarray}
f_0(\bR) &=& f(\bR) \\
\ell^x(\bR) &=& f(\bR) + f(\bR + \hy + \hz) \\
\ell^y(\bR) &=& f(\bR) + f(\bR + \hx + \hz) \\
\ell^z(\bR) &=& f(\bR) + f(\bR + \hx + \hy) \text{,}
\end{eqnarray}
where the notation suggests the connection to the X-cube model.

We define ${\cal S}_f$ to be the subgroup of ${\cal S}_a$ generated by $\{ \pi [f_0(\bR)] \}$, while ${\cal S}_\ell$ is the subgroup generated by $\{ \pi [ \ell^{\mu}(\bR) ] \}$.  It is clear that both ${\cal S}_f$ and ${\cal S}_\ell$ are submodules under the reduced translation symmetry, and ${\cal S}_a = {\cal S}_f \oplus {\cal S}_\ell$ as a direct sum of modules.  What we need to do is establish the isomorphism with the X-cube fracton and lineon fusion theories.  

First we consider
\begin{equation}
\pi[f_0(\bR) ] = P_{yz}(x) + P_{xz}(y) + P_{xy}(z) \text{,}
\end{equation}
where $\bR = (x,y,z)$ and $x,y,z$ are all even.  It is immediately apparent from this expression that ${\cal S}_f$ is isomorphic to the fracton fusion theory of the X-cube model -- the only difference is the restriction that $x,y,z$ are even, but because the reduced translation symmetry is also only for translation vectors with even components, the $\zt$-modules are isomorphic.

Next we consider
\begin{equation}
\pi[\ell^x(\bR) ] = P_{xz}(y) + P_{xz}(y+1) + P_{xy}(z) + P_{xy}(z+1) \text{,}
\end{equation}
where $\bR = (x,y,z)$ and $x,y,z$ are all even, and with similar expressions for $\ell^y$ and $\ell^z$.  From these expressions, given $p \in {\cal S}_\ell$, it is clear that $Q_{yz}(x) = Q_{yz}(x+1)$,
$Q_{xz}(y) = Q_{xz}(y+1)$ and $Q_{xy}(z) = Q_{xy}(z+1)$.  Therefore the coefficients $Q_{\mu \nu}(n)$ with $n$ odd are superfluous, and we obtain an isomorphic $\zt$-module if we replace the generators with
\begin{equation}
\pi[\ell^x(\bR) ] \to P_{xz}(y) +  P_{xy}(z) \text{,}
\end{equation}
with similar expressions for $\ell^y$ and $\ell^z$.  Therefore, following the above argument for the isomorphism of ${\cal S}_f$ with the X-cube  fracton fusion theory, we see that ${\cal S}_\ell$ is isomorphic to the lineon fusion theory of the X-cube model.

\section{Statistical processes from local moves}
\label{sec:statistics}

\subsection{Local moves}
\label{subsec:localmoves}

We have already seen how including translation symmetry in the fusion theory allows us to describe the mobility of excitations, and distinguish fractons, lineons and planons.  Here, we go farther and start from the fusion theory to describe  processes in which excitations move through space.  We will be particularly interested in statistical processes, from which robust phase factors originating from long-range statistical interactions among excitations can be extracted.

We will be interested in processes where an initial configuration of excitations evolves under some adiabatic change of parameters.  In more conventional topological phases, it is enough to start with an initial configuration of excitations that are moved as a function of time, without any excitations being created or destroyed.  In general, moving a fracton or sub-dimensional excitation may require creating other non-trivial excitations, so we will need to consider processes in which excitations are created and destroyed.  It is convenient to abstract away from the language of adiabatic evolution, and to view a process as a sequence of discrete local moves.  Each local move can be realized by acting with a local operator supported on a ball of radius $r_{loc}$, where $r_{loc}$ is some arbitrary but fixed length scale.  In the processes of interest, excitations are moved over large length scales of order $r_{stat}$, and statistical phase factors will only be well-defined in the limit $r_{stat} \to \infty$, keeping $r_{loc}$ fixed.  (Note that we always take $r_{stat} \ll L$, where $L$ is the linear system size.)

To describe local moves, we start with the observation that any set of excitations that fuses to the trivial excitation is locally createable.   For instance, the pair of X-cube model vertex excitations
\begin{equation}
\ell^x(\br) + \ell^x(\br + \hx) = (1 + t_{\hat{x}} ) \ell^x(\br)  \label{eqn:lineon-local-move}
\end{equation}
is locally createable, and this corresponds to the existence of a string operator creating the two excitations of the pair at its endpoints.  Acting with this string operator on an initial state realizes a local move where the two vertex excitations are inserted into the initial state and fused with any existing excitations.  Depending on the initial state, the same local move can create the pair of vertex excitations (if they are not already present), can destroy the pair of vertex excitations (if they are both present in the initial state), or can move a vertex excitation from one position to another (if one of the two excitations is present in the initial state).  Similarly, while no pair of cube excitations in the X-cube model is locally createable, the set of four such excitations given by
\begin{equation}
(1 + t_{\hat{x}} + t_{\hat{y}} + t_{\hat{x} + \hat{y}} ) f(\br) \label{eqn:fracton-local-move}
\end{equation}
is locally createable, and insertion of this set of excitations is thus a local move.

More generally, we would like to describe local moves with reference only to the fusion theory, since this encodes universal properties of the gapped excitations.  Given a superselection sector $s \in {\cal S}$, we are interested in linear combinations of translations satisfying
\begin{equation}
\big( \sum_{\ba} c_{\ba} t_{\ba} \big) s = 0 \text{,} \label{eqn:local-move}
\end{equation}
where the $c_{\ba}$ are integers and only finitely many $c_{\ba}$ are non-zero.  The left-hand side of this equation corresponds to a set of excitations whose relative positions are known, and that collectively fuse to the trivial excitation.  As long as the excitations are not too far apart, these excitations fit into a ball of radius $r_{loc}$, and we have a local move where this set of excitations is inserted into some initial state and fused with any excitations already present.  We can always multiply Eq.~(\ref{eqn:local-move}) by any translation $t_{\ba}$, so this local move can be made anywhere in space.  If $s$ is an element of order $n$ (\emph{i.e.} $n s = 0$ for a positive integer $n$, and $k s \neq 0$ for $0 < k < n$), then we can restrict the coefficients $c_{\ba}$ to run from $0$ to $n-1$, because terms $n t_{\ba}$ in Eq.~(\ref{eqn:local-move}) correspond to trivial excitations that do not need to be included in the description of the local move.  

This discussion can be formalized in a useful way.  First, given $s \in {\cal S}$, we define a ring $R_s$ as follows.  If $s$ is of infinite order, then $R_s = \zt$.  If $s$ is of order $n$, then we take the quotient $R_s = \zt / (n)$, where $(n) \subset \zt$ is the ideal generated by $n$.  The ring $R_s$ is the set of formal linear combinations of translation operations $t_{\ba}$, with coefficients valued in $\z_n$; that is, we take the coefficients $c_{\ba}$ to be valued in $\z_n$.  The reason to focus on $R_s$ rather than $\zt$ is that each non-zero element $r \in R_s$ corresponds to a set of non-trivial excitations obtained by translating $s$, where the non-trivial excitations are simply the the terms in $r s$ obtained by writing $r$ as a linear combination of translations.

Next, we are interested in elements $r \in R_s$ with $r s = 0$. The set of such elements forms an ideal $I_s \subset R_s$, and  each non-zero element $r \in I_s$ corresponds to a set of non-trivial excitations that fuse to the trivial excitation.  As long as these excitations are not too far apart, then non-zero elements $r \in I_s$ correspond to local moves.  Moreover, because $R_s$ is a Noetherian ring, the ideal $I_s$ is finitely generated;\footnote{We thank Jeongwan Haah for pointing this out.} that is, $I_s$ is the set of linear combinations of a set of $k$ generators $\{ q_1, \dots, q_k \}$.  Formally, $I_s = \{ r_1 q_1 + \cdots + r_k q_k | r_i \in R_s \}$, and we write $I_s = (q_1, \dots, q_k)$.  Because there are only a finite number of generators, we can choose $r_{loc}$ so that all the generators correspond to local moves.  It then follows that a general non-zero element of $I_s$ corresponds to a sequence of local moves.  For simplicity, we refer to $I_s$ as the ideal of local moves of $s$.  We note that two sectors related by translation have the same ring $R_s$ and the same ideal of local moves; that is,  $R_s = R_{t_{\ba} s}$ and $I_s = I_{t_{\ba} s}$, as is easily shown.

To illustrate these constructions, we consider the fracton and lineon excitations of the X-cube model.  In the X-cube model, the ideal of local moves $I_f$ for a fracton excitation $f = \pi[ f(\br) ]$ is
\begin{equation}
I_f = (1 + t_{\hx} + t_{\hy} + t_{\hx + \hy}, 1 + t_{\hy} + t_{\hz} + t_{\hy + \hz}, 1 + t_{\hx} + t_{\hz} + t_{\hx + \hz}) \text{.} \nonumber
\end{equation}
These generators correspond to the obvious local moves where we insert four cube excitations at the corners of a plaquette.
For a lineon excitation $\ell^z = \pi[ \ell^z(\br) ]$, we have
\begin{equation}
I_{\ell^z} = (1 + t_{\hz}, 1 + t_{\hx} + t_{\hy} + t_{\hx + \hy}) \text{,} \nonumber
\end{equation}
with corresponding expressions for $\ell^x$ and $\ell^y$ lineons.

For the most part, in this paper we will consider local moves obtained as above using the translation symmetry.  However, in general, there are local moves not contained in the ideal of local moves of any excitation.  As an example, we consider the fracton fusion theory of the X-cube model upon breaking the $t_{\hx}$ symmetry, but preserving the translation subgroup generated by $t_{2 \hx}$, $t_{\hy}$ and $t_{\hz}$.  We let $f_1 = \pi[ f(\br) ]$ and $f_2 = t_{\hx} f_1 = \pi[ f(\br + \hx) ]$.  $f_1$ and $f_2$ are generators for ${\cal S}_b$ with the lower translation symmetry.  Unlike with the full translation symmetry, $f_1$ and $f_2$ are not related by translation.  The set of excitations $\{ f_1, f_2, t_{\hz} f_1, t_{\hz} f_2 \}$ is locally createable, and cannot be obtained from any ideal of local moves, but certainly corresponds to a legitimate local move as long as $r_{loc}$ is chosen  large enough to accommodate all the excitations in a ball of radius $r_{loc}$.  This can always be done because, even though we have no information about the relative location of two excitations in sectors $f_1$ and $f_2$, two such excitations will be a finite distance apart.

\subsection{Statistical processes}
\label{subsec:statproc}

Now we describe statistical processes in terms of local moves, illustrating our general discussion with a trivial -- or, more properly, very familiar -- example from the $d=2$ toric code model.  Statistical processes of some fracton phases are discussed in Sec.~\ref{sec:statfracton}.  There are some similarities to an earlier approach to statistics in $d=2$ \cite{levin03fermions}.

A statistical process acts on a specified initial configuration of well-separated excitations, which is assumed to be the same as the final configuration at the end of the process, so that the system returns to its initial state multiplied by a phase factor.  To specify a configuration of excitations, we need to know the superselection sector and the spatial position of each excitation.  The process is defined in terms of $n$ steps, each of which is a sequence of a large number of local moves.  The number of moves in each step is large because we are interested in processes that move excitations over distances on the order of $r_{stat}$, which we refer to as ``large'' distances, in contrast to ``small'' distances on the order of $r_{loc}$.  It is important to keep in mind that excitations can be created and destroyed during a step.

\begin{figure}
    \centering
    \begin{subfigure}{0.45\columnwidth}
        \includegraphics[width=\textwidth]{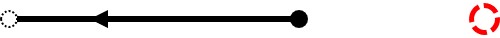}
        \caption{Step 1}
    \end{subfigure}
    \hfill \vrule \hfill
    \begin{subfigure}{0.45\columnwidth}
        \includegraphics[width=\textwidth]{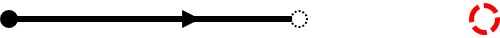}
        \caption{Step 2}
    \end{subfigure}
    \\ \hrule 
    \vspace{0.5cm}
    \begin{subfigure}{0.5\columnwidth}
        \includegraphics[width=\textwidth]{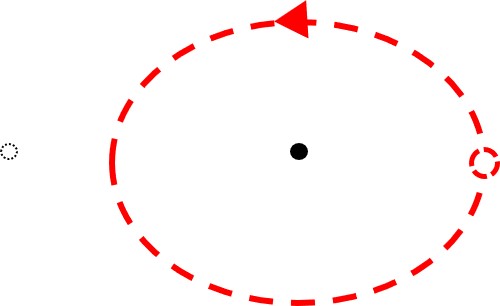}
        \caption{Step 3}
    \end{subfigure}
    \caption{Braiding of an $e$ particle (red dashed circle) around a $m$ particle (black solid circle) in the $d=2$ toric code.  First the $m$ particle is moved to the right from its initial position, then the $e$ particle is braided around it, and finally the $m$ particle is moved back to its initial position.  Each step is sequence of a large number of local moves.}\label{fig:toricCode_process}
\end{figure}

A familiar example of a statistical process is the braiding of an $e$ particle (vertex excitation) around an $m$ particle (plaquette excitation) in the $d=2$ toric code.  Fig.~\ref{fig:toricCode_process} illustrates this process in three steps.  It is important that these steps can be rearranged and executed in two different orders, resulting in two different processes, one in which $e$ braids around $m$ (123; as shown), and one in which no braiding occurs (132).  Looking only at the process where the braiding occurs is not enough to extract the statistical phase factor, because there are also \emph{local} contributions to the phase that accumulate during each step.  However, these local contributions are the same for both processes, and the statistical phase factor (which is $-1$ in this case) is the relative phase between the two processes.

In this toric code example, in order to have a well-defined separation between local and non-local contributions to the phase, it is important that the excitation being moved during each step stays a large distance away from the other excitation.  We can generalize this requirement to an arbitrary statistical process, by first defining the spatial support of each step.  Each local move is implemented by acting with an operator supported in a small spatial region, and taking the union of these regions for the local moves in a step gives the spatial support of the step.  We then require that the support of each step is a large distance away from all excitations except those that are moved, created or destroyed by the step.  We refer to this as the \emph{spatial separation requirement}, and it is sufficient to obtain a well-defined statistical phase factor by comparing two different arrangements of the steps in a process.  It may be possible to weaken the spatial separation requirement in an interesting way, but we will not pursue this here.

Any statistical phase factor we find should be a property of an equivalence class of statistical processes; that is, the statistical phase should not depend on all the details of a particular process, and there should be a notion of small deformations of processes that do not affect the statistical phase.  As a familiar illustration of this idea, we can visualize the toric code example in terms of world lines for the $e$ and $m$ excitations, where the $m$ world line executes a full braid around the $e$ world line as a function of time.  We can obtain equivalent statistical processes by smoothly deforming the world lines, forbidding crossings.  While this notion of equivalence invokes a continuum picture, we can formulate essentially the same idea in the discrete language of local moves appropriate for our treatment.

In general, we allow two types of small deformations of a statistical process.  For type I deformations, among the sequence of local moves that make up a step, we replace any sub-sequence of local moves supported in a small region with an equivalent sub-sequence, also of small support.  By equivalent sub-sequence, we mean that the two sub-sequences have the same effect on the initial configuration of excitations before the sub-sequence.  It is allowed to replace the empty sub-sequence (\emph{i.e.} the sub-sequence with no moves) with a sub-sequence of small support that does not create any excitations.  It is important that the new process obtained by this deformation has to satisfy the spatial separation requirement.  In type II deformations, we  add or remove a local move at the end of a step.  This changes the configuration of excitations after the step, and a corresponding move must be added or removed at the beginning of some other step (not necessarily the next step) so that the final and initial conditions match for each pair of adjacent steps.  Again, the new process has to satisfy the spatial separation requirement.  We can also contemplate other kinds of equivalence operations that might be important in a more systematic treatment, such as splitting steps into two steps and joining adjacent steps together, but we leave this for future work as it will not be needed for the present discussion.  Along the same lines, we believe there is nothing fundamental about working with steps, and it should be possible to formulate a description of statistical processes as (long) sequences of local moves.

\begin{figure}
    \centering
    \begin{subfigure}{\columnwidth}
        \includegraphics[width=\textwidth]{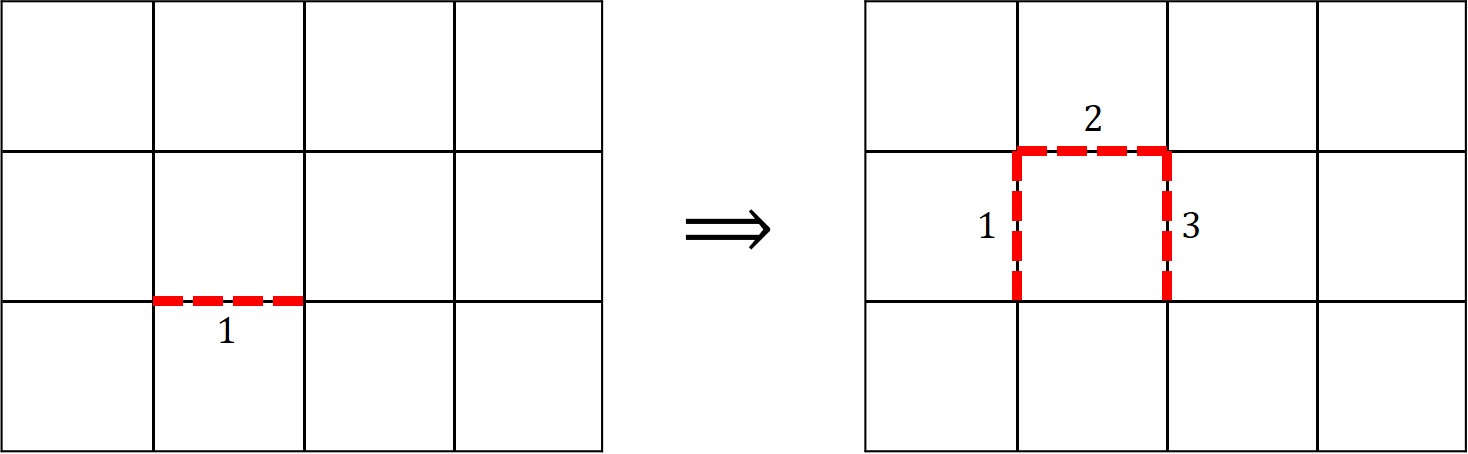}
        \caption{\label{fig:TCdef_a}}
    \end{subfigure}
     \\%add desired spacing between images, e. g. ~, \quad, \qquad, \hfill etc. 
    \begin{subfigure}{\columnwidth}
        \includegraphics[width=\textwidth]{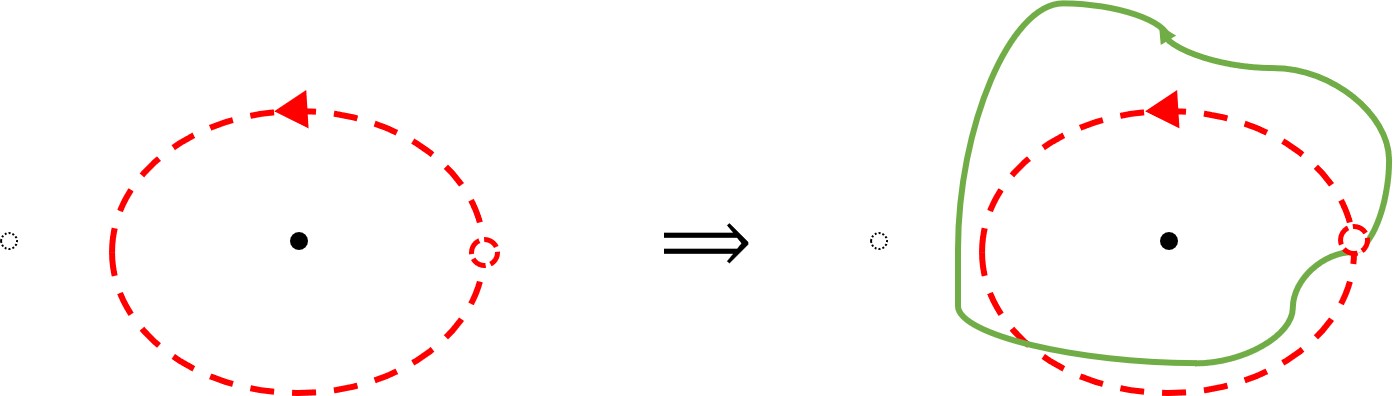}
                \caption{\label{fig:TCdef_b}}
    \end{subfigure}
   \\
     \begin{subfigure}{\columnwidth}
        \includegraphics[width=\textwidth]{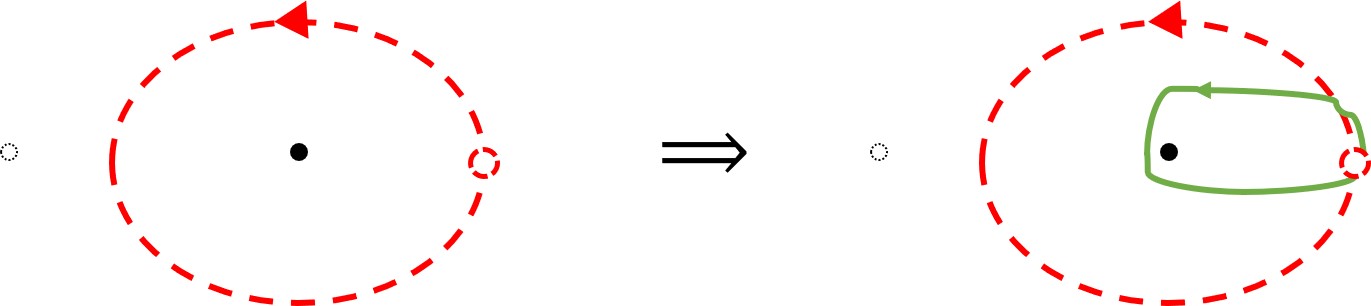}
                \caption{\label{fig:TCdef_c}}
    \end{subfigure}
    \\
    \begin{subfigure}{\columnwidth}
        \includegraphics[width=\textwidth]{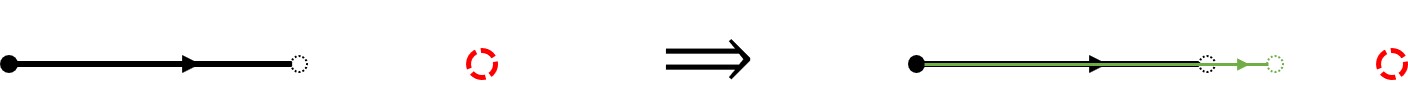}
                \caption{\label{fig:TCdef_d}}
    \end{subfigure}
    \caption{Deformations of the $e$-$m$ braiding process in the $d=2$ toric code.  The panels of the figure are described in the text.}
    \label{fig:TCdef}
\end{figure}

We illustrate this discussion using the toric code example.  Individual local moves correspond to inserting a pair of $e$ ($m$) particles on adjacent sites (plaquettes).  Fig.~\ref{fig:TCdef_a} shows a type I deformation where a single such local move (dashed red link) is replaced with a sequence of three local moves.  Fig.~\ref{fig:TCdef_b} shows an allowed deformation (solid green line) of step 2 in the statistical processes that can be obtained as a sequence of small type I deformations.  Fig.~\ref{fig:TCdef_c} shows a deformation (solid green line) that is not allowed because it violates the spatial separation requirement, with the $e$ particle coming too close to the $m$ particle.  The spatial separation requirement implies that the particle world lines cannot cross.  Finally, Fig.~\ref{fig:TCdef_d} shows a deformation that modifies step 1 by moving the final location of the $m$ particle, which can be obtained as a sequence of small type II deformations, and for which step 3 must also be modified accordingly.

We note that in all the statistical processes we consider in this paper, only a finite number of local moves distinct up to an overall translation appear, which means it is always possible to choose a fixed $r_{loc}$ accommodating all the local moves.

\section{Statistical processes of some fracton phases}
\label{sec:statfracton}

\subsection{X-cube and semionic X-cube model: fracton-lineon process}
\label{subsec:fracton-lineon}

\begin{figure}[b]
  \includegraphics[scale=0.45]{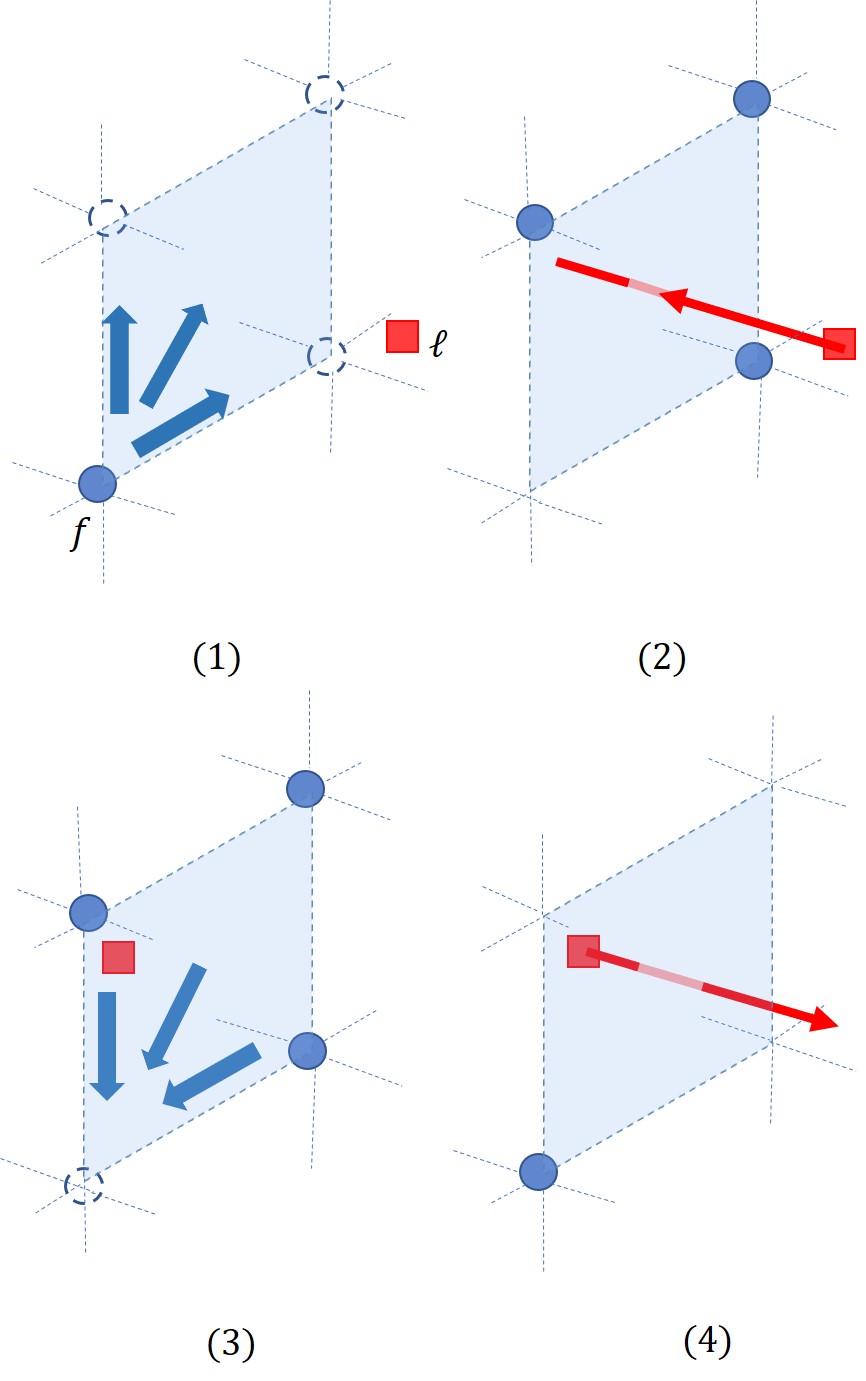}
 \caption{Fracton-lineon process in the X-cube model.  In step (1) the fracton cube excitation $f$ (blue circle) is destroyed, and three cube excitations are created at the other corners of the blue membrane.  In step (2), the lineon vertex excitation $\ell$ (red square) is moved through the blue membrane.  Step (3) then undoes step (1), and step (4) undoes step (2).}
 \label{fig:FLprocess}
 \end{figure}
 
We now discuss some examples of statistical processes of fracton phases, beginning with the ``fracton-lineon'' process of the X-cube fracton phase illustrated in Fig.~\ref{fig:FLprocess}.  Because the semionic X-cube and ordinary X-cube models have the same fusion theory, this is also a statistical process in the semionic X-cube model.  The process consists of four steps as shown, with an initial configuration of a single cube excitation $f$ and a single vertex excitation $\ell$.  Steps 1 and 3 are sequences of the local moves described by Eq.~(\ref{eqn:fracton-local-move}), and steps 2 and 4 are built from the local moves expressed in Eq.~(\ref{eqn:lineon-local-move}).  Two orderings of the steps are possible; one is $1234$ as shown in Fig.~\ref{fig:FLprocess}, and the other is $1324$.  The latter process is obviously deformable to the trivial process where the particles do not move, while in the former fracton-lineon process there is a kind of ``braiding'' between the lineon and the fracton associated with the lineon's string piercing the membrane in step 2.  Each step is associated with a product of Pauli operators in the X-cube model, and the phase accumulated during the process can be obtained by multiplying these operators in the order given.  It is easy to see that we obtain a phase of $-1$ for the ordering $1234$, and a phase of $1$ for the $1324$ ordering, so there is a non-trivial statistical phase of $-1$ associated with the fracton-lineon process.  We emphasize that the non-trivial statistical phase is an invariant characterizing deformation classes of process, and by computing it we prove that the fracton-lineon process cannot be deformed to the trivial process.
 
\begin{figure}[t]
  \includegraphics[width=\columnwidth]{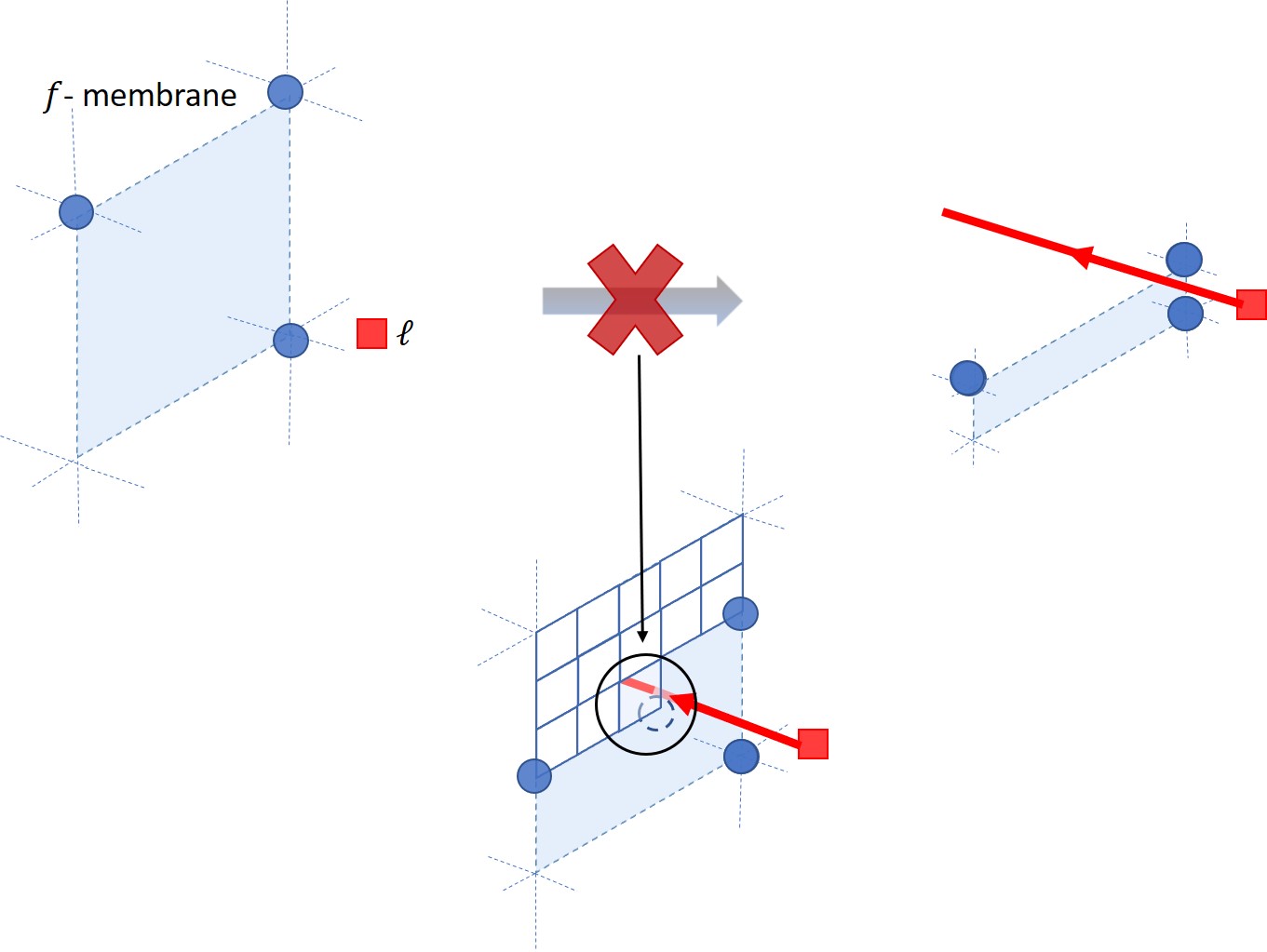}
 \caption{The spatial separation requirement is violated if steps 1 and 3 of the fracton-lineon process are deformed by removing local moves to shrink the membrane to that illustrated on the top-right.  At an intermediate stage of the deformation (bottom), a cube excitation is present within the black circle after step 1, and the lineon passes nearby this cube excitation during step 2, violating the spatial separation requirement.}
 \label{fig:SSviolation}
 \end{figure}

It is interesting to understand the obstruction that arises if we try to deform the fracton-lineon process to the trivial process.  For instance, we might like to reduce the size of the membrane appearing in steps 1 and 3, so that the lineon string no longer passes through it.  Such an attempted deformation is illustrated Fig.~\ref{fig:SSviolation}, from which we see that the obstruction is provided by the spatial separation requirement.

\begin{figure}[t]
  \includegraphics[width=\columnwidth]{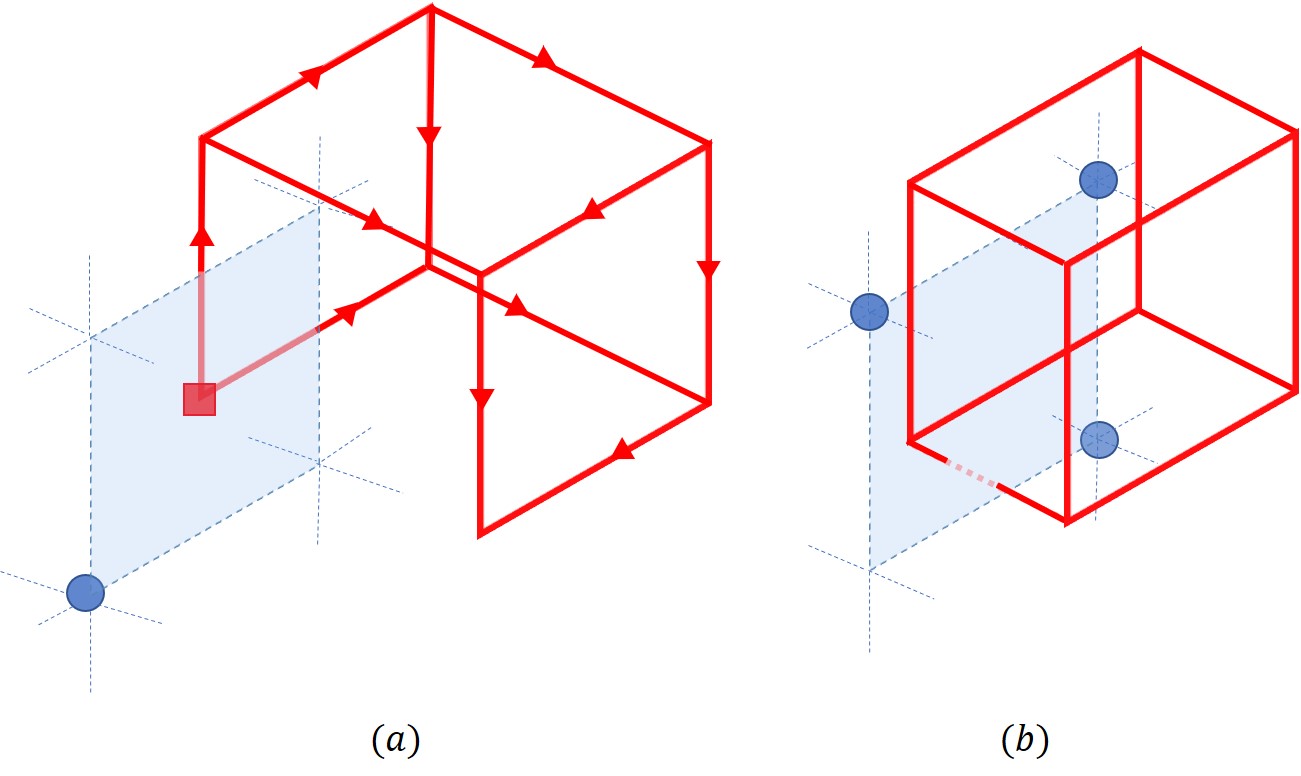}
 \caption{(a) An allowed deformation of step 4 in the fracton-lineon process.  (b) Step 2 in a modified fracton-lineon process where step 4 has been deformed away, as described in the text.}
 \label{fig:step4modified}
 \end{figure}

The fracton-lineon process can be deformed in interesting ways.  For instance, step 4 can be deformed as shown in Fig.~\ref{fig:step4modified}(a).  This deformation relies on the fact that $\ell^x(\br) + \ell^y(\br) + \ell^z(\br)  = 0$, so while a single lineon cannot ``turn a corner,'' it can split into a pair of lineons moving in perpendicular directions.  It is important to note that step 2 cannot be similarly deformed without violating the spatial separation requirement.  Once step 4 is deformed as in Fig.~\ref{fig:step4modified}(a), the fracton-lineon process can be further deformed, by deleting moves from the beginning of step 4 and adding them to the end of step 2, until step 4 becomes trivial.  This results in a process where step 2 is as shown in Fig.~\ref{fig:step4modified}(b), with a lineon ``cage'' enclosing the fracton at the upper-right corner.  This process can be viewed as the remote detection of the enclosed fracton via the lineon process (or its corresponding operator) illustrated in Fig.~\ref{fig:step4modified}(b).  Such remote detection has been discussed previously in other works \cite{ma17coupled,slagle17robust,shirley2019fractional,bulmash18braiding}, and is analogous to the remote detection of a $d=2$ toric code $m$ particle by braiding an $e$ particle around it, as in step 2 of Fig.~\ref{fig:toricCode_process}. 

\begin{figure}[b]
  \includegraphics[width=0.7\columnwidth]{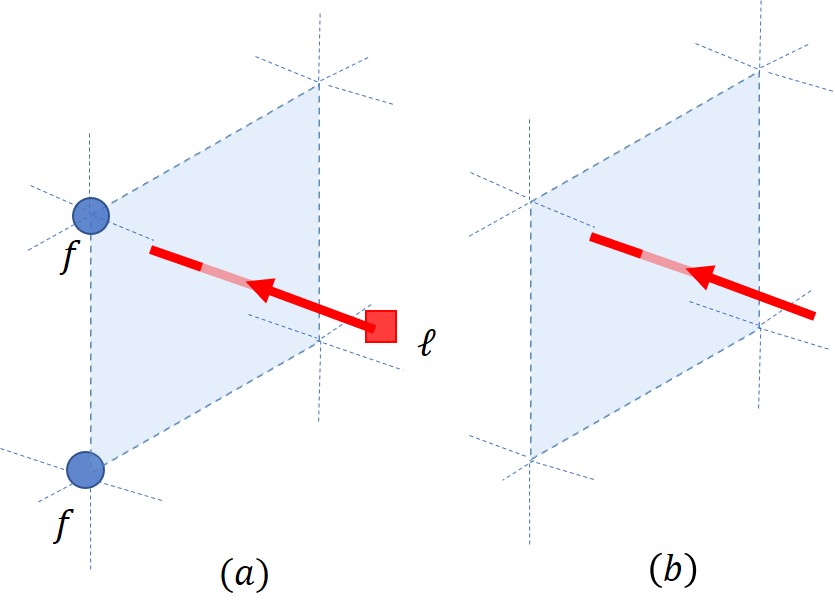}
 \caption{Two processes related to the fracton-lineon process by changing the initial state, but with the same sequence of steps.  (a) shows the initial state in the lineon-planon process, which consists of a pair of $f$ fractons that together form a planon, and the same lineon $\ell$ as in the fracton-lineon process. (b)  shows the initial state for a vacuum process where no particles are present.}
 \label{fig:related}
 \end{figure}
 
 \begin{figure*}
  \includegraphics[width=0.8\textwidth]{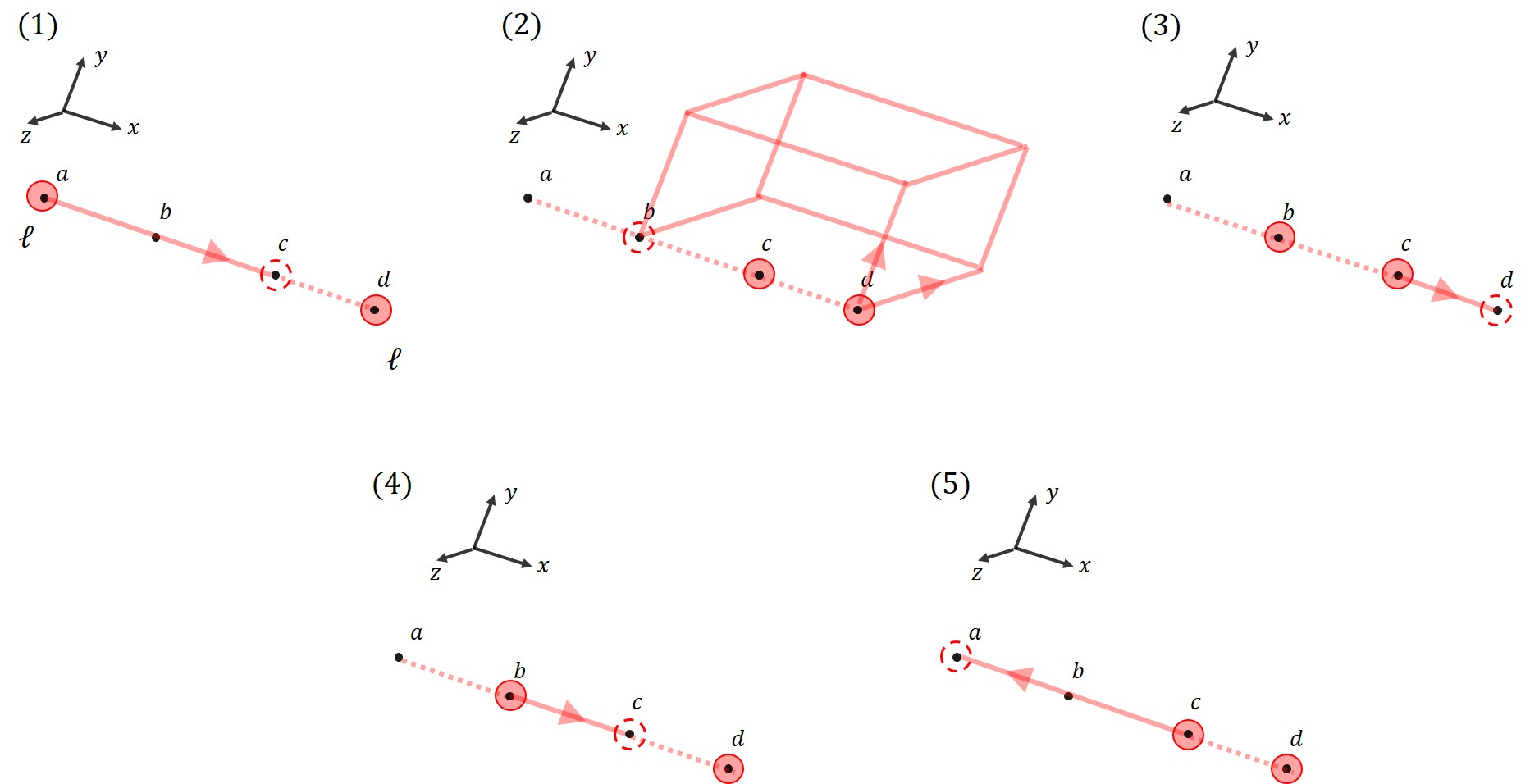}
 \caption{Lineon-lineon exchange process in the X-cube model.  The initial configuration consists of two vertex excitations $\ell^x(0)$ and $\ell^x(x_d \hat{x})$ on the $x$-axis at positions labeled $a$ and $d$, respectively, with $x_d \hat{x}$ the position vector of $d$ and $x_d \in \z$.  In step 1, the lineon at $a$ is moved to position $c$.  Step 2 is the most non-trivial step:  the lineon initially at $d$ is moved off the $x$-axis by splitting it into $\ell^y$ and $\ell^z$ vertex excitations, producing three $\ell^x$ excitations off the axis, which are then moved in the $-x$-direction and eventually recombined into a $\ell^x$ excitation at position $b$.  In the remaining three steps, $\ell^x$ excitations are moved along the $x$-axis as shown.  If the steps are executed in the order shown, then the two lineons are exchanged during this process.  However, if instead the steps are carried out in the order $15243$, no exchange occurs.}
 \label{fig:lineon-lineon}
 \end{figure*}

We also note that there are a number of processes related to the fracton-lineon process by choosing a different initial state, but with precisely the same steps.  For instance, Fig.~\ref{fig:related}(a) corresponds to a process where the planon composed of two cube excitations is braided with the lineon $\ell$ \cite{vijay16fracton}.  Similarly, there is a ``vacuum process'' [Fig.~\ref{fig:related}(b)] where no particles are present in the initial state.  In this process, for instance, step 1 creates a set of four fractons out of the vacuum.  These processes  have the same $-1$ statistical phase factor as for the fracton-lineon process, because in each case one multiplies precisely the same products of operators to extract the statistical phase.

For Abelian anyons in $d=2$, it is well known that the fusion theory puts constraints on the possible statistical phase factors that can be associated with a given braiding process.  For instance, $+1$ and $-1$ are the only two statistical phases consistent with fusion in the $e$-$m$ braiding process of the toric code (Fig.~\ref{fig:toricCode_process}).  This can be seen by noting that a composite of two $e$ particles fuses to the trivial sector, and so must braid trivially with $m$, while observing that the phase factor for such a process should be a product of two $e$-$m$ braiding phase factors.  More formally, the latter observation is the bilinearity property of the braiding phase factor.  Precisely the same argument shows that for any system where the fusion theory is the same as that in the X-cube model, the only possible statistical phase factors for the fracton-lineon process are $+1$ or $-1$.  Indeed, the semionic X-cube model has the same fusion theory and the same $-1$ statistical phase factor for its fracton-lineon process.  It is tempting to try to formalize this discussion by defining a bilinear form on the $\zt$-module of the fusion theory, and while this may be sensible it will require more thought in future work, because composites of cube excitations and vertex excitations have different mobility and cannot all undergo the same fracton-lineon process. It is important to emphasize that there is a difference between attaching statistical phase factors to processes, which clearly makes sense, and attaching them to sets of elements in the fusion theory, which makes sense for anyons in $d=2$ but is not necessarily sensible in fracton phases.

\subsection{X-cube and semionic X-cube model: lineon-lineon process}
\label{subsec:lineon-lineon}

There is also a lineon-lineon exchange process in the X-cube and semionic X-cube models, in which two vertex excitations mobile along the same line are exchanged, as shown in Fig.~\ref{fig:lineon-lineon}.  One might guess that the exchange statistics is not well-defined because the particles are confined to the same line, and they cannot be exchanged without getting close together. However this expectation is too na\"{\i}ve, because one of the lineons can be moved out of the way by splitting it into three excitations away from the other's line of motion, as illustrated in step 2 of Fig.~\ref{fig:lineon-lineon}.  Such a process was discussed previously in Ref.~\onlinecite{song18twisted}.  Other related processes discussed in prior work are mentioned briefly at the end of this section.

This process is easily seen to give a trivial statistical phase factor for the vertex excitations in the X-cube model, as their string operators are simply products of $X$ Pauli operators.  However, a non-trivial statistical phase occurs for other lineon excitations in the X-cube model.  For instance, we can replace the initial-state excitations $\ell^x(0)$ and $\ell^x(x_d \hat{x})$ by
\begin{equation}
\tilde{\ell}^x(0) = \ell^x(0) + f( \hat{c} ) + f( \hat{c} - \hat{z} ) \text{,}
\end{equation}
and
\begin{equation}
\tilde{\ell}^x(x_d \hat{x}) = \ell^x(x_d \hat{x}) + f( x_d \hx + \hat{c} ) + f( x_d \hx + \hat{c} - \hz ) \text{,}
\end{equation}
respectively.  To avoid confusion, here we use a single coordinate system to label positions of vertex and cube excitations, and $\hat{c} = (\hat{x} + \hat{y} +\hat{z}) /2$ is the offset between the simple cubic lattices of vertices and cube centers.  These new excitations are composites of vertex excitations with planons composed of two cube excitations, where the planons are mobile in the $xy$-plane.

It is convenient to view the new excitations in terms of the coupled-layer construction (see Refs.~\onlinecite{ma17coupled,vijay17coupled} and Sec.~\ref{sec:xcube-pstring}), which lets us describe the excitations of the X-cube model in terms of excitations of $d=2$ toric code layers.  This will allow us to compute the statistical phase factor in a very simple manner.  Schematically (\emph{i.e.} dropping position labels), we can write for both these excitations
\begin{equation}
\tilde{\ell}^x = e^y + e^z + m^z  = e^y + \epsilon^z \text{,}
\end{equation}
where $e^{\mu}$, $m^{\mu}$ and $\epsilon^{\mu}$ are the $d=2$ toric code excitations in a layer with normal in the $\mu$-direction.  This follows from the fact that the vertex excitations are bound states of toric-code $e$-particles in perpendicular layers, and the composite of two cube excitations is identified with a toric-code $m$-particle.  We also replace the $\ell^y$ and $\ell^z$ excitations that appear (as string operators) in step 2, according to $\ell^{y,z} \to \tilde{\ell}^{y,z}$, with $\tilde{\ell}^y = \ell^y + m^z$ and $\tilde{\ell}^z = \ell^z$.  The criterion for choosing these replacements is that the same statistical process  must still be possible for the new excitations.

Now, each step in Fig.~\ref{fig:lineon-lineon} can be viewed as a process where the $d=2$ toric code excitations are moved within their corresponding planes.  In step 2 some pairs of $d=2$ excitations are created, moved, and then annihilated.  It is straightforward to see that for the $12345$ ordering of steps, the $d=2$ particles making up the initial-state $\tilde{\ell}^x$ excitations are exchanged, \emph{i.e.} there is an exchange of $e^y$ with $e^y$ and an exchange of $\epsilon^z$ with $\epsilon^z$.  These exchanges are counterclockwise in the $xz$ and $xy$ planes, respectively.  On the other hand, for the $15243$ ordering, no particles are exchanged.  Therefore the statistical phase is $-1$ and comes from the $\epsilon^z-\epsilon^z$ exchange.

Exchanges of $\tilde{\ell}^y$ and $\tilde{\ell}^z$ are also well-defined statistical processes, obtained by rotating Fig.~\ref{fig:lineon-lineon}.  We denote the statistical phase angles for these exchanges by $\theta^{x,y,z}_{\ell \ell}$, and we have $\theta^x_{\ell \ell} = \theta^y_{\ell \ell} = \pi$ while $\theta^z_{\ell \ell} = 0$.  These statistical phase angles are defined for \emph{any} triple of excitations $\tilde{\ell}^{x,y,z}$ satisfying $t_{\hat{\mu}} \tilde{\ell}^{\mu} = \tilde{\ell}^{\mu}$ (where $\hat{\mu}$ is the unit vector in the $\mu$-direction) and $\sum_{\mu = x,y,z}  \tilde{\ell}^{\mu} = 0$.  We observe that 
\begin{equation}
\sum_{\mu = x,y,z} \theta^{\mu}_{\ell \ell}  = 0 \mod 2\pi  \text{,}
\end{equation}
both for the triple $\tilde{\ell}^{x,y,z}$ of vertex excitations bound to planons, and for the triple $\ell^{x,y,z}$ simply consisting of vertex excitations.

We  also consider this process for the semionic X-cube models, where the corresponding $d=2$ excitations are the bosonic flux $m^{\mu}$, the semion $s^{\mu}$, and the anti-semion ${s'}^{\mu} = s^{\mu} + m^{\mu}$.  The vertex excitations are composites of semions in two perpendicular layers, \emph{e.g.} $\ell^x = s^y + s^z$.  For the triple of vertex excitations $\ell^x, \ell^y$ and $\ell^z$, we have $\theta^{\mu}_{\ell \ell} = \pi$, and
\begin{equation}
\sum_{\mu = x,y,z} \theta^{\mu}_{\ell \ell}  = \pi \mod 2\pi  \text{,}
\end{equation}
which suggests there is a fundamental difference between the semionic X-cube model and the ordinary X-cube model.  We discuss this further in Sec.~\ref{sec:2Xcube}, where we prove that these models realize distinct translation-invariant fracton phases.

Other processes involving lineons have been discussed previously.  The mutual statistics of two lineons moving along perpendicular, intersecting lines was discussed in Ref.~\onlinecite{ma17coupled} and shown to be non-trivial for the vertex excitations of the semionic X-cube model.  It appears likely that this ``full braid'' lineon process is related to the lineon-lineon exchange process discussed here.  Also, Ref.~\onlinecite{yizhi18twisted}, in analogy with the ribbon twists that can be used to diagnose the topological spin of anyons in $d=2$, considered adding a twist to a lineon string operator, and referred to this as a ``boxing process.''  The relationship of this twist to a statistical process in the sense we define it here is not obvious, but it appears likely that our lineon-lineon statistical process corresponds to the twist of Ref.~\onlinecite{yizhi18twisted}, in the same sense that anyon exchange corresponds to a ribbon twist in $d=2$.

\subsection{Checkerboard model:  fracton-fracton process}
\label{subsec:checkerboard-fracton-fracton}

We also discuss a process involving fractons in the checkerboard model, that does not appear to have a simple interpreration as a braiding or exchange of mobile particles.  Of course, giving up some translation symmetry, the checkerboard model is equivalent to two copies of the X-cube model, so there should be an interpretation of this process in terms of X-cube model statistical processes, perhaps some combination of fracton-lineon and lineon-lineon processes.  However, it is not clear that this viewpoint provides a simple interpretation of the process in question.

\begin{figure}[b]
  \includegraphics[width=\columnwidth]{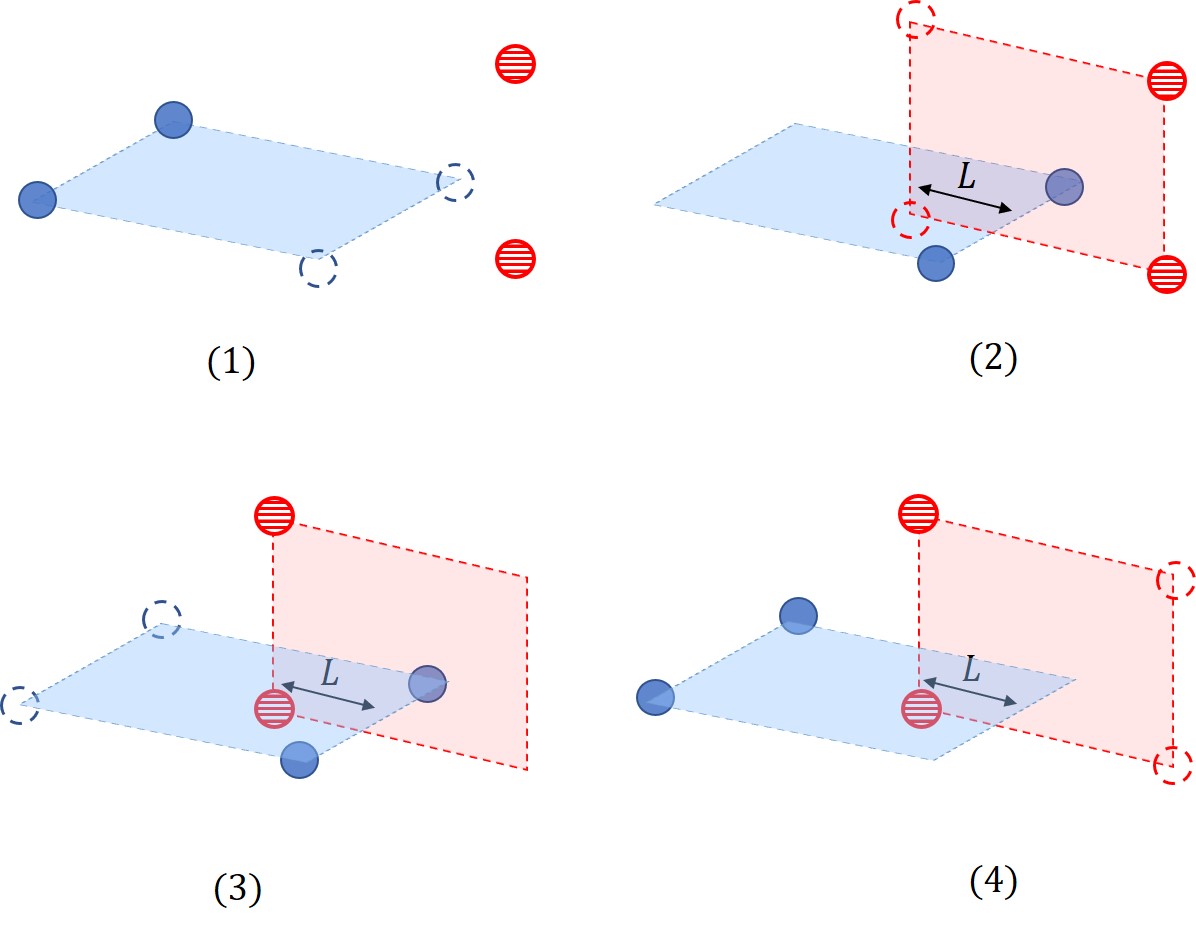}
 \caption{Fracton-fracton process in the checkerboard model.  The initial configuration is shown in (1) and has two pairs of fractons, one pair separated along the $x$-axis and the other pair separated along the $z$-axis.   The blue solid circle fractons are excitations of the $A_c$ terms in the checkerboard Hamiltonian, while the red  striped circle fractons are excitations of the $B_c$ terms.   In step (1), the blue solid circle pair of fractons is moved to the positions indicated by the dashed blue circles.  Step (2) proceeds similarly for the red striped circle fractons.  Carrying out this process on the lattice, the two membrane operators used to implement steps (1) and (2) intersect on a line segment containing $L$ lattice sites.  Step (3) reverses step (1), and step (4) reverses step (2).  Multiplying together the operators implementing each step gives a phase of $(-1)^L$ for the order of steps shown, but a trivial phase for the order $1324$.}
 \label{fig:checkerboard-process}
 \end{figure}
 
The fracton-fracton process is illustrated in Fig.~\ref{fig:checkerboard-process} and described in the caption.  As noted there, the process is characterized by an integer length $L$, and results in a statistical phase of $(-1)^L$.  Evidently, changing $L$ by one gives two inequivalent processes.  While this is somewhat surprising, the $(-1)^L$ is clearly a statistical phase factor and proves that there are at least two deformation classes of such processes in the fusion theory of the checkerboard model, depending on the parity of $L$.  It is straightforward to consider simple deformations that would change $L$ and see that the spatial separation requirement is violated in the course of the deformation.

\section{X-cube and semionic X-cube models realize different fracton phases}
\label{sec:2Xcube}

As an application of the theory of fusion and statistics, we prove that the X-cube and semionic X-cube models realize distinct translation-invariant fracton phases.  Indeed, this was already argued to be the case in Ref.~\onlinecite{ma17coupled} where the semionic X-cube model was introduced.  The argument was based on the observation that the lineon-lineon mutual statistics in one model could not be transformed into that of the other by binding planons.  However, without a clear framework to characterize fusion and statistics this argument did not definitively establish the difference between these two phases.  Indeed this distinction was called into question by Ref.~\onlinecite{shirley2019fractional}, which showed that the X-cube and semionic X-cube models realize the same foliated fracton phase.  (See Sec.~\ref{sec:intro} for a brief discussion of the notion of foliated fracton phase.)

Here, we prove that these models realize distinct translation-invariant fracton phases.  Notably, this conclusion holds even if we break the translation symmetry down to any subgroup isomorphic to $\z^3$, \emph{i.e.} if the translation-breaking corresponds to an enlargement of the crystalline unit cell.  One may also ask what happens if  translation symmetry is broken completely, but our approach does not apply in that case, and we are not aware of any means currently available to provide a sharp answer.  Nevertheless, the distinction between these two phases should not be thought of as a fragile one that disappears upon breaking translation symmetry.

First of all, as noted in Sec.~\ref{sec:xcube-pstring}, the fusion theories for the X-cube and semionic X-cube models are isomorphic, so any distinction must come from statistical processes.  We will consider the lineon-lineon process described in Sec.~\ref{subsec:lineon-lineon}, which is sufficient to establish the desired result.  We let ${\cal S}_{Xc}$ and ${\cal S}_{sXc}$ be the fusion theories for the X-cube and semionic X-cube models, respectively.  Even though there is no difference between these two $\zt$-modules, it is better for our purposes to view them as isomorphic  and not to identify them.  This allows us to describe additional structure attached to  ${\cal S}_{Xc}$ and to ${\cal S}_{sXc}$ that partly characterizes the statistics in the two models.  ${\cal S}_{Xc}$ is generated by the cube excitation $f(0)$ and vertex excitations $\ell^x(0)$, $\ell^y(0)$.  In this section we abuse notation and denote elements of ${\cal S}_{Xc}$ by representative elements of the excitation group ${\cal E}$, without explicitly writing the projection map $\pi$.  We denote the corresponding cube and vertex excitations for ${\cal S}_{sXc}$ by $f_s(0)$, $\ell_s^x(0)$, $\ell_s^y(0)$.  There is a $\zt$-module isomorphism 
$\alpha : {\cal S}_{Xc} \to {\cal S}_{sXc}$ defined by mapping the generators of ${\cal S}_{Xc}$ to the corresponding generators of ${\cal S}_{sXc}$ in the obvious way.  It is important to note that while $\alpha$ is an isomorphism of $\zt$-modules, it does not preserve statistics, \emph{i.e.} it does not preserve the statistical phase factors attached to statistical processes.

The approach will be to assume the two models are in the same phase and obtain a contradiction.   Under this assumption, there is an isomorphism of $\zt$-modules $\beta : {\cal S}_{sXc} \to {\cal S}_{Xc}$ that, unlike $\alpha$, does preserve statistics.  Therefore there is an automorphism $\gamma = \beta \circ \alpha$, with $\gamma : {\cal S}_{Xc} \to {\cal S}_{Xc}$, such that the statistical properties of $\gamma[\ell^x(0)], \gamma[ \ell^y (0) ]$ and $\gamma[ \ell^z (0) ]$ are identical to those of $\ell_s^x(0)$, $\ell_s^y(0)$ and $\ell_s^z(0)$.  In particular, for the triple of excitations $\gamma[\ell^x(0)], \gamma[ \ell^y (0) ], \gamma[ \ell^z (0) ]$, we must have the lineon-lineon statistical phase factors $\theta^{\mu}_{\ell \ell} = \pi$, and moreover $\sum_{\mu = x,y,z} \theta^{\mu}_{\ell \ell} = \pi \mod 2\pi$.

We will show that no such automorphism $\gamma$ exists.  We let $\tilde{\ell}^{\mu} \equiv  \gamma[ \ell^\mu (0) ]$.  Focusing first on $\tilde{\ell}^x$, we have in general
\begin{equation}
\tilde{\ell}^x = \sum_{\br} a(\br) f(\br) + \sum_{\br, \mu} b_{\mu}(\br) \ell^{\mu}(\br) \text{,} \label{eqn:tlx-gen}
\end{equation}
where the first sum is over cube centers, the second sum over vertex centers, and the coefficients $a(\br)$ and $b_{\mu}(\br)$ take values in $\{0, 1\}$.  Because $\gamma$ is an automorphism of $\zt$-modules we have $t_{\hx} \tilde{\ell}^x = \tilde{\ell}^x$.  This implies that the first term in Eq.~(\ref{eqn:tlx-gen}) reduces to a sum of $m^y$ and $m^z$ planons, while the second term is a sum of $\ell^x(\br)$ excitations.  These conclusions are easily reached using the description of the fusion theory in terms of plane charges.  It is convenient to represent the $\ell^x(\br)$ excitations in terms of toric code $e^{\mu}$ particles, by $\ell^x(x,y,z) = e^y(y) + e^z(z)$.  Formally we can write
\begin{eqnarray}
\tilde{\ell}^x &=& \sum_{y \in \z} [  \alpha_y(y) m^y(y) + \beta_y(y) e^y(y) ] \nonumber  \\
 &+& \sum_{z \in \z} [ \alpha_z(z) m^z(z) + \beta_z(z) e^z(z) ] \text{.}
\end{eqnarray}
with coefficients $\alpha_{\mu}, \beta_{\mu} \in \{ 0,1 \}$, and similarly we have
\begin{eqnarray}
\tilde{\ell}^y &=& \sum_{x \in \z} [  \alpha_x(x) m^x(x) + \beta_x(x) e^x(x) ] \nonumber  \\
 &+& \sum_{z \in \z} [ \alpha_z(z) m^z(z) + \beta_z(z) e^z(z) ]  \\
 \tilde{\ell}^z &=& \sum_{x \in \z} [  \alpha_x(x) m^x(x) + \beta_x(x) e^x(x) ] \nonumber  \\
 &+& \sum_{y \in \z} [ \alpha_y(y) m^y(y) + \beta_y(y) e^y(y) ] \text{.}  
 \end{eqnarray}
 Here, the same $\alpha_{\mu}$ and $\beta_{\mu}$ coefficients appear in the three expressions due to the requirement that $\sum_{\mu} \tilde{\ell}^{\mu} = 0$.

Now we consider the lineon-lineon process and compute $\theta^{\mu}_{\ell \ell}$ for the triple of excitations $\tilde{\ell}^{\mu}$.  First we consider $\theta^x_{\ell \ell}$.  During this process, each excitation appearing in the sum over $y$ is exchanged with another identical such excitation, giving a contribution of $\pi$ to the statistical phase when $\alpha_y(y) = \beta_y(y) = 1$ [in this case the excitation appearing is $\epsilon^y(y)$], and a contribution of zero otherwise.   The corresponding statement holds for the sum over $z$, and the total statistical phase is thus
\begin{equation}
\theta^x_{\ell \ell} = \Big[ \pi \sum_{j \in \z} \big( \alpha_y(j) \beta_y(j) + \alpha_z(j) \beta_z(j) \big) \Big] \mod 2\pi \text{.}
\end{equation}
The obvious corresponding expressions hold for $\theta^y_{\ell \ell}$ and $\theta^z_{\ell \ell}$, and therefore we have
\begin{equation}
\sum_{\mu = x,y,z} \theta^{\mu}_{\ell \ell} = \Big[ 2 \pi \sum_{j \in \z} \sum_{\mu = x,y,z}  \alpha_\mu (j) \beta_\mu (j)  \Big] \mod 2\pi = 0 \text{.}
\end{equation}
This is a contradiction, and completes our proof that the ordinary and semionic X-cube models realize distinct translation-invariant fracton phases.

Now we consider breaking translation down to a subgroup isomorphic to $\z^3$, and show that the conclusion is unchanged.  Such a subgroup is generated by three linearly independent translation vectors $\ba_1, \ba_2$ and $\ba_3$, which are integer linear combinations of $\hx, \hy$ and $\hz$.  First we consider $\ba_1 = n_x \hx$, $\ba_2 = n_y \hy$ and $\ba_3 = n_z \hz$, in which case the argument proceeds essentially unchanged.  The key observation is that $t_{n_{\mu} \hat{\mu}} \tilde{\ell}^{\mu} = \tilde{\ell}^{\mu}$ puts exactly the same restriction on the form of $\tilde{\ell}^{\mu}$ as in the case of full translation symmetry.  Next, in the general case, the vectors $N_x \hx, N_y \hy$ and $N_z \hz$ can be obtained as integer linear combinations of $\ba_1, \ba_2$ and $\ba_3$ for some non-zero integers $N_{x,y,z}$.  To see this, we consider $\sum_i n_i \ba_i = N_x \hx$ as the matrix equation
\begin{equation}
\left( \begin{array}{ccc}
a_{1x} & a_{2x} & a_{3x} \\
a_{1y} & a_{2y} & a_{3y} \\
a_{1z} & a_{2z} & a_{3z} 
\end{array}\right)
\left( \begin{array}{c}
n_1 \\ n_2 \\ n_3 
\end{array}\right)
=
\left( \begin{array}{c}
N_x \\ 0 \\ 0
\end{array}\right) \text{.}
\end{equation}
We assume $N_x$ is a non-zero integer, and the equation can be solved for $n_1, n_2, n_3$ by inverting the matrix.  While the $n_i$ thus obtained need not be integers, they are rational numbers because the inverse matrix has rational entries.  Changing $N_x$ by multiplying it by a suitable integer then results in a set of integers $N_x \neq 0$ and $n_1, n_2, n_3$ satisfying the equation.  This shows that the translation group generated by $\ba_1, \ba_2, \ba_3$ has a subgroup generated by $N_x \hx, N_y \hy, N_z \hz$, and using only this subgroup we are back to the first case again.

\section{Fusion of electric and magnetic excitations in ${\rm U}(1)$ tensor gauge theories}
\label{sec:u1fusion}

Here, we develop a fusion theory for gapped electric and magnetic excitations in the deconfined phase of some ${\rm U}(1)$ tensor gauge theories with fracton and sub-dimensional excitations.  There are some differences between these theories and gapped fracton phases.  First of all, there are gapless photon excitations.  Second, even in the deconfined phase, there are sometimes gapped electric or magnetic excitations whose energy cost diverges with the separation from other excitations.  For instance, in the $d=3$ rank-2 scalar charge theory, the energy to create an isolated electric charge grows linearly with the separation from other charges, a phenomenon dubbed electrostatic confinement \cite{pretko17subdimensional}.  Nonetheless, we still expect that superselection sectors and their fusion make sense.  Indeed, this is certainly the case for gapped electric and magnetic charges in ordinary vector ${\rm U}(1)$ gauge theories in $d=3$, and as in that case the presence  gapless photon modes is not expected to present any issues.  Electrostatically confined excitations with diverging energy cost may seem problematic, but, in the examples we consider, the energy density in the electric and magnetic fields still goes to zero far away from the excitation in question.  As pointed out in Ref.~\onlinecite{pretko17subdimensional}, such an isolated electrostatically confined excitation is thus still expected to be stable.  However, in rank-3 and higher tensor gauge theories, there are excitations where the energy density itself diverges at long distances, and these excitations are not stable \cite{pretko17subdimensional}.  We will not consider such examples, for which the notion of superselection sectors does not make sense.  But we do note that our analysis can be applied to tensor gauge theories with such unstable excitations, as long as one restricts attention to the stable excitations whose energy density vanishes in the far field.

Even though gapless photons are present, we do expect that some notion of statistics of gapped finite-energy excitations makes sense.  For instance, in ordinary vector ${\rm U}(1)$ gauge theories it is well known that binding a bosonic electric charge with a bosonic magnetic monopole results in a fermionic dyon \cite{goldhaber76connection}.  However, we leave the exploration of such phenomena in tensor gauge theories for future work.

We treat two different lattice rank-2 scalar charge theories here, both in $d=3$.  These theories belong to a family of theories labeled by $(m,n)$, where $m$ and $n$ are positive relatively prime integers, which were characterized by considering the Higgs mechanism whereby a charge-2 matter field condenses, and studying the gapped phase that arises \cite{bulmash18Higgs}. In contrast, we provide a characterization that applies directly to the deconfined phase of the $(m,n)$ theory, which we illustrate with two examples.  First we discuss the $(1,1)$ theory, sometimes referred to simply as \emph{the} scalar charge theory, and for which the mobility of electrically charged excitations was first discussed in Ref.~\onlinecite{pretko17subdimensional}.  In Sec.~\ref{sec:21scalarcharge}, we obtain the fusion theory for the electric excitations of the $(2,1)$ theory.  We find that the $(1,1)$ and $(2,1)$ theories have different fusion properties, and in particular the $(2,1)$ theory has additional conservation laws not present in the $(1,1)$ theory.  We see that these theories thus describe different gapless fracton phases, even though their physics in the photon sector is identical.  This is consistent with the results of Ref.~\onlinecite{bulmash18Higgs}, where it was shown that the $(1,1)$ theory becomes four copies of the $d=3$ $\zz$ toric code upon condensing charge-2 matter (a result also obtained in Ref.~\onlinecite{ma18Higgs}), while the $(2,1)$ theory enters the X-cube fracton phase.

\subsection{Fusion in the $(1,1)$ scalar charge theory}
\label{sec:11scalarcharge}

We consider a family of  rank-2 ${\rm U}(1)$ tensor gauge theories on the simple cubic lattice, following Ref.~\onlinecite{bulmash18Higgs} (see also Ref.~\onlinecite{ma18Higgs} for a discussion of one of these theories).  We briefly introduce these models here; for more details the reader is referred to Refs.~\onlinecite{ma18Higgs, bulmash18Higgs}.  The degrees of freedom are quantum rotor variables that realize a symmetric rank-2 ${\rm U}(1)$ tensor gauge field on the lattice, with the electric field operators $E_{\mu \nu}$ taking integer eigenvalues, and the gauge potential $A_{\mu \nu}$ a $2\pi$-periodic phase field.    The diagonal components of $E$ and $A$ reside on lattice sites, while the off-diagonal components live on square plaquettes, and $[ A_{\mu \nu} , E_{\mu \nu} ] = -i$ for variables on the same site.  We also introduce charged matter fields on lattice sites, with $n_{\br}$ taking integer eigenvalues, and $\theta_{\br}$ the corresponding canonically conjugate $2\pi$-periodic phase.  We consider the Gauss law
\begin{equation}
\begin{aligned}
& m ( \Delta_x \Delta_x E_{xx} +\Delta_y \Delta_y E_{yy} + \Delta_z \Delta_z E_{zz}) \\ &+ n (\Delta_x  \Delta_y E_{xy} + \Delta_y  \Delta_z E_{yz} +\Delta_x  \Delta_z E_{xz}   )   = n_{\bf r} \text{,}  \label{eqn:lattice-gauss}
\end{aligned}
\end{equation}
where $m$ and $n$ are positive relatively prime integers.

We introduce the magnetic field tensor $B_{\mu \nu}$, which is traceless but not symmetric, whose diagonal (off-diagonal) components reside on the sites (plaquettes) of the dual cubic lattice, and which is given by
\begin{equation}
B_{\mu \nu} = \left\{
\begin{array}{ll}
\epsilon_{\mu \lambda \sigma} \Delta_{\lambda} A_{\sigma \nu} , &  \mu = \nu \\
\epsilon_{\mu \nu \lambda} ( m \Delta_{\nu} A_{\lambda \nu} - n \Delta_{\lambda} A_{\nu \nu} ) , & \mu \neq \nu 
\end{array}\right. \text{,}
\end{equation}
where in the second line there are no implied summations and $\lambda \neq \mu, \nu$.
The Hamiltonian is
\begin{eqnarray}
H &=&  U \sum_{{\bf r},\mu \leq \nu} E^2_{\mu\nu} - K \sum_{{\bf r},\mu,\nu} \cos ( B_{\mu\nu} ) \nonumber \\
&+& u \sum_{\bf r} n^2_{\bf r}  - J \sum_{{\bf r},\mu < \nu} \cos \biggl[ n \Delta_{\mu} \Delta_{\nu} \theta -A_{\mu \nu} \biggr] \nonumber  \\
&-& J \sum_{{\bf r},\mu} \cos \biggl[m  \Delta_{\mu} \Delta_{\mu} \theta -A_{\mu \mu} \biggr] \text{,}
\end{eqnarray}
where $U, K, u, J > 0$.  When $u \gg J$ and $K$ is large, there is a deconfined phase with gapped electric and magnetic charges, and a gapless photon excitation with five polarizations, whose low-energy Gaussian description can be obtained by expanding the $- K \cos (B)$ term to quadratic order.  Within this Gaussian theory, the integers $m$ and $n$ play no role and can be removed by a rescaling.  

We now specialize to the $(1,1)$ theory, and first develop the fusion theory of the gapped electric charge excitations.  Configurations of electric charge are elements of the $\zt$-module ${\cal E}_e = \bigoplus_{\br} \z$, where the direct sum is over all sites of the cubic lattice.  We denote the generator of the $\z$ summand at $\br$ by $f(\br)$, so that $e \in {\cal E}_e$ can be written $e = \sum_{\br} q(\br) f(\br)$, where $q(\br) \in \z$ and only finitely many $q(\br)$ are non-zero.  Translation acts in the obvious way by $t_{\ba} f(\br) = f(\br + \ba)$.  Any locally createable charge configuration can be created by acting with a finite product of exponentials ${\exp[ i (A_{\mu \nu} - \Delta_{\mu} \Delta_{\nu} \theta) ]}$.  Single such operators create the generators of ${\cal L}_e \subset {\cal E}_e$, which are ${2 f(\br) - f(\br - \hz) - f(\br + \hz)}$, ${f(\br) - f(\br + \hx) - f(\br + \hy) + f(\br + \hx + \hy)}$, and the elements obtained from these by acting with cubic symmetry operations.  These charge configurations can be understood as quadrupoles on a line of three sites, and on the four corners of a square, respectively.

To proceed, we will define a $\zt$-module ${\cal S}_e$, and then show it is isomorphic to ${\cal E}_e / {\cal L}_e$.  It is known that the total charge and dipole moment of an electrically charged excitation are conserved \cite{pretko17subdimensional}, in the sense that these quantities cannot be changed by any local process, so we may expect that different superselection sectors should be labeled by their charge and dipole moment.  We thus define ${\cal S}_e = \z \oplus \z \oplus \z \oplus \z$, and write an element $s \in {\cal S}_e$ by $s = Q \alpha + \boldsymbol{D} \cdot \boldsymbol{\beta}$, where $\alpha$, $\beta_x$, $\beta_y$ and $\beta_z$ generate the four $\z$ summands, and $Q, D_{\mu} \in \z$.  We will see that $Q$ corresponds to an excitation's charge and $\boldsymbol{D}$ to its dipole moment.  We define translation to act on ${\cal S}_e$ by
\begin{eqnarray}
t_{\ba} \alpha &=& \alpha + \ba \cdot \boldsymbol{\beta} \\
t_{\ba} \beta_{\mu} &=& \beta_{\mu} \text{.}
\end{eqnarray}
This gives the desired transformations of charge and dipole moment, namely $Q \to Q$ and $\boldsymbol{D} \to \boldsymbol{D} + Q \ba$.  It is clear that excitations with $Q \neq 0$ are fractons, while $Q  = 0$ excitations are fully mobile.

To establish an isomorphism between ${\cal S}_e$ and ${\cal E}_e / {\cal L}_e$, we first define a map $\pi : {\cal E}_e \to {\cal S}_e$ by
\begin{equation}
\pi [ f(\br) ] = \alpha + \br \cdot \boldsymbol{\beta} \text{,}
\end{equation}
which is easily checked to be a map of $\zt$-modules, and which simply expresses mathematically that a charge $q(\br) = 1$ contributes $1$ to the total charge and $\br$ to the total dipole moment.    It is easy to see that ${\cal L}_e \subset \operatorname{ker} \pi$, so we have an induced map $\pi_S : {\cal E}_e / {\cal L}_e \to {\cal S}_e$.  In Appendix~\ref{app:11electric}, we show that $\pi_S$ is an isomorphism.  We thus see that charge and dipole moment give a complete labeling of different electric  superselection sectors in the $(1,1)$ scalar charge theory.  Electrically charged excitations in this theory do not have any additional conserved quantities.  This is in contrast to other $(m,n)$ theories, which do have additional conservation laws as we illustrate in Sec.~\ref{sec:21scalarcharge} for the $(2,1)$ theory.

Now we turn to the gapped magnetically charged excitations of the $(1,1)$ scalar charge theory.  It is convenient to suppress electric charge excitations completely and pass to a dual description, where the magnetic charge is an operator $n_{\mu}(\br)$ with integer eigenvalues, on the links of the dual cubic lattice \cite{ma18Higgs}, and we can view the charge as a vector quantity.  The pair $(\br, \mu)$ denotes the link directed in the $+ \mu$ direction from the dual site $\br$, where $\mu = x,y,z$. In this dual description the magnetic field tensor $B_{\mu \nu}$  has real eigenvalues, is again traceless but not symmetric, and the diagonal (off-diagonal) components reside on sites (plaquettes) of the dual cubic lattice.  We have the dual Gauss law
\begin{equation}
\Delta_{\mu} B_{\mu \nu} = n_{\nu}(\br) \text{.}
\end{equation}
As noted in Ref.~\onlinecite{ma18Higgs}, using the Gauss law and integrating by parts, one can show that the vector charge $Q_{\mu} = \sum_{\br} n_{\mu}(\br)$ and its scalar first moment $P = \sum_{\br, \mu} r_{\mu} n_{\mu}(\br)$ are both conserved.

We introduce the $\zt$-module ${\cal E}_m$ whose elements are configurations of magnetic charge, and ${\cal E}_m = \bigoplus_{\br, \mu} \z$, where the direct sum is over links of the dual cubic lattice.   We denote the generator of the $\z$ summand for the link at $(\br, \mu)$ by $g(\br, \mu)$, so a general element $e \in {\cal E}_m$ is expressed
\begin{equation}
e = \sum_{\br, \mu} q(\br, \mu) g(\br, \mu) \text{,}
\end{equation}
with $q(\br, \mu) \in \z$.  Translation acts in the obvious way by $t_{\ba} g(\br, \mu) = g(\br + \ba, \mu)$.

\begin{figure}
  \includegraphics[width=0.9\columnwidth]{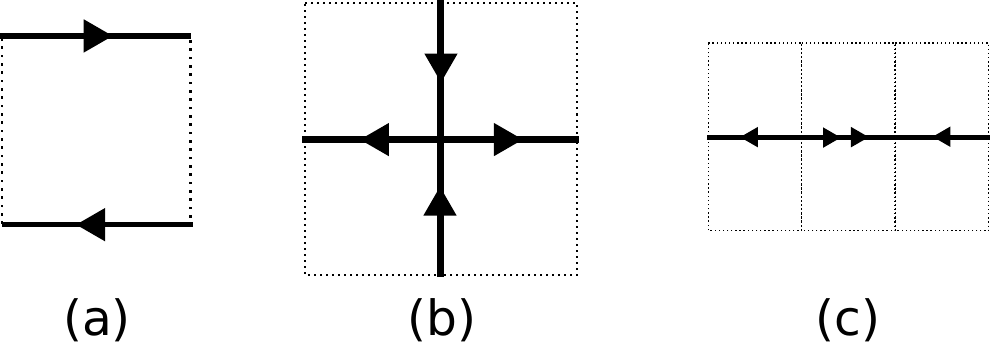}
 \caption{(a) and (b) depict the generators of ${\cal L}_m \subset {\cal E}_m$, the $\zt$-module of locally createable magnetically charged excitations of the $(1,1)$ scalar charge theory.  Solid lines represent $q(\br,\mu) = \pm 1$, with the sign given by the direction of the arrow, while dotted lines represent $q(\br, \mu) = 0$.  (c) Shows an element of ${\cal L}_m$ that can be obtained by adding generators; elements of this form are used in Appendix~\ref{app:11magnetic} in a proof that $\operatorname{ker} \pi \subset {\cal L}_m$.  The line with a double arrow represents $q(\br, \mu) = \pm 2$.}
  \label{fig:Lm_generators}
 \end{figure}

Letting $\phi_\mu$ be the phase field canonically conjugate to $n_{\mu}$, and $\alpha_{\mu \nu}$ the dual gauge potential conjugate to $B_{\mu \nu}$, any locally createable configuration of magnetic charge can be created by acting with a finite product of operators $\exp[i (\alpha_{\mu \nu} - \Delta_{\mu} \phi_{\nu} )]$ for $\mu \neq \nu$, and also ${\exp[i (\alpha_{\mu \mu} - \alpha_{\nu \nu} - \Delta_{\mu} \phi_{\nu} + \Delta_{\nu} \phi_{\nu} )]}$.  These operators correspond to generators of ${\cal L}_m \subset {\cal E}_m$ of the form ${g(\br + \hy, x) - g(\br, x)}$ [Fig.~\ref{fig:Lm_generators}(a)] and 
${g(\br, x) - g(\br - \hx, x) - g(\br, y) + g(\br-\hy,y)}$ [Fig.~\ref{fig:Lm_generators}(b)], and others obtained by acting on these with cubic symmetry operations.

As for the electric sector, we define a $\zt$-module ${\cal S}_m$, and then show it is isomorphic to ${\cal E}_m / {\cal L}_m$.  As a group, ${\cal S}_m = \z \oplus \z \oplus \z \oplus \z$, with generators $\alpha_x, \alpha_y, \alpha_z$ and $\beta$ for the four $\z$ summands.  A general element $s \in {\cal S}_m$ can be written $s = \boldsymbol{Q} \cdot \boldsymbol{\alpha} + P \beta$, where we will see that $Q_{\mu}, P \in \z$ correspond to the total vector charge and moment $P$.  Translation is defined by act by
\begin{eqnarray}
t_{\ba} \alpha_{\mu} &=& \alpha_{\mu} + a_{\mu} \beta \\
t_{\ba} \beta &=& \beta \text{,}
\end{eqnarray}
which gives the expected transformations $\boldsymbol{Q} \to \boldsymbol{Q}$ and $P \to P + \boldsymbol{Q} \cdot \ba$.  Clearly excitations with $\boldsymbol{Q} = 0$ are fully mobile, while excitations with $\boldsymbol{Q} \neq 0$ are planons mobile in the plane perpendicular to $\boldsymbol{Q}$.

We define a map $\pi : {\cal E}_m \to {\cal S}_m$ by
\begin{equation}
\pi [ g(\br, \mu) ] = \alpha_{\mu} + r_{\mu} \beta \text{,}
\end{equation}
which is easily checked to be a map of $\zt$-modules.  Moreover, the generators of ${\cal L}_m$ map to zero under $\pi$, so ${\cal L}_m \subset \operatorname{ker} \pi$, and $\pi$ induces a map $\pi_S : {\cal E}_m / {\cal L}_m \to {\cal S}_m$, which in Appendix~\ref{app:11magnetic} is shown to be an isomorphism, so that ${\cal S}_m$ is the fusion theory of the magnetically charged excitations of the $(1,1)$ scalar charge theory.

\subsection{Fusion in the $(2,1)$ scalar charge theory}
\label{sec:21scalarcharge}

Now we consider the $(2,1)$ scalar charge theory, which is defined above in Sec.~\ref{sec:11scalarcharge}.  We will restrict our attention to the fusion theory of the gapped electric charge excitations.  Electric charge configurations are the same $\zt$-module ${\cal E}_e$ defined above.  Here, the generators of ${\cal L}_e$ are ${2[2 f(\br) - f(\br - \hz) - f(\br + \hz)}]$ (where the overall factor is $m=2$) and ${f(\br) - f(\br + \hx) - f(\br + \hy) + f(\br + \hx + \hy)}$, and elements related to these by cubic symmetry.  

In this case, we define a $\zt$-module $\Sigma_e$, and then we will show that a certain submodule ${\cal S}_e \subset \Sigma_e$, which we will characterize, is isomorphic to ${\cal E}_e / {\cal L}_e$.  We define $\Sigma_e$ to be the direct sum of modules $\Sigma_e = {\cal S}_e^{(1,1)} \oplus {\cal P}_b$, where ${\cal S}_e^{(1,1)}$ is the fusion theory of electric charge excitations in the $(1,1)$ scalar charge theory, and where ${\cal P}_b \subset {\cal P}$ is the fracton fusion theory of the X-cube model and ${\cal P}$ is the $\zt$-module of $\zz$ plane charges, as described in Sec.~\ref{subsubsec:Xcube-fracton}.  We recall that an element $p \in {\cal P}$ can be characterized by integers $k_p, l_p, m_p$, and $p \in {\cal P}_b$ if and only if $k_p, l_p$ and $m_p$ are either all odd or all even.  We can also further label elements $p \in {\cal P}_b$ by integers $k^{o,e}_p, l^{o,e}_p, m^{o,e}_p$ introduced in Sec.~\ref{sec:checkerboard-fusion}.  

Now we define the obvious map $\pi : {\cal E}_e \to \Sigma_e$ by its value on generators:
\begin{equation}
\pi [ f(\br) ] = \alpha + r_{\mu} \beta_{\mu} + P_{yz}(x) + P_{xz}(y) + P_{xy}(z) \text{,}
\end{equation}
where $\br = (r_x, r_y, r_z) = (x,y,z)$.

Our first task is to show that $\operatorname{ker} \pi = {\cal L}_e$, so that $\pi_S : {\cal E} / {\cal L}_e \to \Sigma_e$ is injective, and we can identify ${\cal S}_e = \operatorname{im} \pi_S$.  It is clear that ${\cal L}_e \subset \operatorname{ker} \pi$; we need to show the reverse inclusion.  We consider $e \in {\cal E}_e$ with $\pi(e) = 0$.  This implies that the charge and dipole moment of $e$ are zero, so $e$ can be written as a linear combination of generators of ${\cal L}_e^{(1,1)}$, the locally createable electric charge configurations of the $(1,1)$ theory.  Starting with this expansion and subtracting off generators of ${\cal L}_e$, we get a new element
\begin{equation}
e' = \sum_{\br, \mu} n_{\mu}(\br) \Omega_{\mu}(\br)  \text{,}
\end{equation}
which satisfies $\pi(e') = 0$, where the coefficients $n_{\mu}(\br) \in \{0,1\}$, and where
\begin{equation}
\Omega_{\mu}(\br) = f(\br + \hat{\mu}) - 2 f(\br) + f(\br - \hat{\mu}) \text{.}
\end{equation}
We have for instance
\begin{equation}
\pi [ \Omega_{z}(\br) ] = P_{xy}(r_z  +1) + P_{xy}(r_z - 1) \text{.}
\end{equation}
Since only a finite number of the $n_{\mu}(\br)$ are non-zero, it is thus clear that we must have $n_{\mu}(\br) = 0$ for all $\br, \mu$,  in order for all the plane charges in $\pi(e')$ to vanish, so that $e' = 0$.  We have thus shown $e \in {\cal L}_e$.

We now characterize ${\cal S}_e \subset \Sigma_e$.  For any $e \in {\cal E}_e$, we write
\begin{equation}
\pi(e) = Q \alpha + \boldsymbol{D} \cdot \boldsymbol{\beta} + p \text{,}
\end{equation}
where $p \in {\cal P}_b$.  It is easy to check that 
\begin{eqnarray}
Q \operatorname{mod} 2 &=& k_p \operatorname{mod} 2 = l_p \operatorname{mod} 2 = m_p \operatorname{mod} 2  \label{eqn:21c1} \\
D_x \operatorname{mod} 2 &=& k_p^o \operatorname{mod} 2 \\
D_y \operatorname{mod} 2 &=& l_p^o \operatorname{mod} 2 \\
D_z \operatorname{mod} 2 &=& m_p^o \operatorname{mod} 2 \text{.} \label{eqn:21c4}
\end{eqnarray}

We have shown that if $s \in {\cal S}_e$, then these constraints are satisfied.  In fact, the converse is also true, \emph{i.e.} if these contraints hold, then $s \in {\cal S}_e$, so Eqs.~(\ref{eqn:21c1}-\ref{eqn:21c4}) give a full characterization of ${\cal S}_e$.  To show this, we let 
\begin{equation}
e = Q f(0) + \sum_{\mu = x,y,z} D_{\mu} [ f(\hat{\mu}) - f(0)]  +  \sum_{\br, \mu} n_{\mu}(\br) \Omega_{\mu}(\br) 
\text{,}
\end{equation}
for arbitrary integers $Q$ and $D_{\mu}$, and with arbitrary $n_{\mu}(\br) \in \{0,1\}$, and observe that
\begin{equation}
\pi(e) = Q \alpha + \boldsymbol{D} \cdot \boldsymbol{\beta} + p \text{.}
\end{equation}
By choosing $n_{\mu}(\br)$, we can obtain any $p \in {\cal P}_b$ satisfying the constraints Eqs.~(\ref{eqn:21c1}-\ref{eqn:21c4}).

\section{Discussion}
\label{sec:discussion}

In this work, we introduced a set of ideas to describe the fusion and statistics of gapped excitations in Abelian fracton phases.  Because statistical processes involve motion of excitations, the mobility of excitations needs to be encoded in any such theory.  Our approach accomplishes this by assuming lattice translation symmetry, and our results can thus be viewed as a characterization of translation-invariant fracton phases.  Translation symmetry is incorporated into the fusion theory as an action on superselection sectors, which results in the mathematical structure of a $\zt$-module, which we computed for some gapped type I fracton models and ${\rm U}(1)$ symmetric-tensor gauge theories.  This leads to a description of statistical processes in terms of local moves, and we discussed some examples of statistical processes in the X-cube and checkerboard fracton models. As an application of our approach, we gave a proof that the X-cube and semionic X-cube models realize distinct translation-invariant fracton phases, even when the translation symmetry is broken down to any subgroup isomorphic to $\z^3$.

Our work opens up a number of questions in the study of fracton matter.  This paper focuses on type I fracton phases, but it would be interesting to consider extensions of our ideas to type II fracton models such as Haah's cubic code.  Our approach to fusion of excitations applies equally well in type II systems.  However, an obstacle to studying type II fracton phases using these methods is a lack of simple characterizations of the fusion theory, along the lines of its description in terms of plane charges for the X-cube model.  If such a description exists, it will be important to find it in future work.  Moreover, it is not obvious to what extent statistical processes are well defined in type II fracton models.  The issue is that in such systems, a process creating an isolated fracton requires surmounting energy barriers that grow logarithmically with its distance from other excitations, due to creation of other excitations at intermediate stages of the process   \cite{brayi11landscape}.  We expect that it should still be possible to identify processes associated with well-defined statistical phase factors, as long as the intermediate-stage excitations can be kept far away from others during the process, but some care will be required.

For anyons in $d=2$, any statistical process can be decomposed as a sequence of elementary exchange and full-braid operations.  Moreover, in conventional iTO phases we are used to the idea that the fusion theory puts constraints on statistics.  An example is the fact that an Abelian excitation in $d=2$ that fuses with itself to vacuum can only be a boson, a fermion, or a semion. These issues are not understood for statistics in fracton phases, and they will need to be addressed in order to develop a more complete theory in the future.  For instance, given the fusion theory for the X-cube model, can any statistical process be decomposed into fracton-lineon and lineon-lineon processes?  How can we attach a complete set of data describing statistics to the fusion theory, and what are the possibilities?

Another direction to explore will be fusion and statistics of non-Abelian fracton phases, of which there are now a few examples \cite{vijay17nonabelian,song18twisted,prem18cagenet, bulmash2019gauging,prem2019gauging}.  It seems reasonable that a fusion theory might be provided by a fusion category with some kind of action of translation symmetry.  If such a mathematical description can be obtained, then it will be interesting to use it as a starting point for a theory of statistical processes of non-Abelian fracton and sub-dimensional excitations.  Ultimately one might like to develop an ``algebraic theory of fractons'' encoding both fusion and statistics in some algebraic structure, analogous to a unitary modular tensor category for anyons in $d=2$.

We close with some general remarks on the nature of fracton phases of matter, which are highlighted by the contrast between the translation-invariant fracton phases we focus on here, and the notion of foliated fracton phases introduced in Refs.~\onlinecite{shirley18manifolds,shirley19entanglement,shirley2019fractional,shirley18checkerboard,shirley18subsystem,slagle18foliated} and briefly described in Sec.~\ref{sec:intro}.  There are (at least) two definitions we can imagine giving for the notion of quantum phases of matter.  One definition is in terms of an equivalence relation defined by adiabatic continuity and the operation of adding ``trivial'' degrees of freedom in a product state.  This definition underlies the translation-invariant fracton phases we discuss.  A second definition is provided by viewing phases of matter as renormalization group (RG) fixed points, which is closer to the point of view taken in work on foliated fracton phases.

Our experience with more conventional quantum phases of matter, including iTO phases, leads to an expectation that these two definitions should agree.  However, the definitions disagree dramatically in fracton phases.  For instance, in order to view the ground state of the X-cube model as an RG fixed point, it is necessary to allow for adding and removing $d=2$ topologically ordered layers \cite{shirley18manifolds}, and this leads to a coarser equivalence relation than in the first definition above.  Nonetheless, one should not simply discard the first definition; in doing so, for instance, one would lose valuable distinctions among states that cannot be adiabatically connected by varying parameters within some particular lattice model (without adding degrees of freedom).  We believe that exploring the tension between these two definitions of phases of matter will be an interesting direction for future work on fracton phases of matter, with potential implications beyond the realm of fractons.

\acknowledgments{MH is grateful to Xie Chen, Jeongwan Haah, Sung-Sik Lee and Dominic Williamson for useful discussions.  SP is grateful to Michael Pretko, Han Ma, Abhinav Prem, and Albert Schmitz for insightful discussions. This research is supported by the U.S. Department of Energy, Office of Science, Basic Energy Sciences (BES) under Award number DE-SC0014415.  This work was begun in fall 2016 at the Kavli Institute for Theoretical Physics, which is supported by the National Science Foundation under Grant No. NSF PHY11-25915.  Some of this work was performed at the Aspen Center for Physics in summer 2018, which is supported by National Science Foundation grant PHY-1607611.}

\appendix

\section{Technical details in fusion theory computations}
\label{app:technical}

In this Appendix we give some details that are skipped in the computation of fusion theories in Sec.~\ref{sec:examples} and Sec.~\ref{sec:11scalarcharge}.

\subsection{Fracton fusion theory of the X-cube model}
\label{app:Xcube-fracton}

First, we show that $\operatorname{ker} \pi \subset {\cal L}_b$, which implies that $\operatorname{ker} \pi = {\cal L}_b$ as claimed in Sec.~\ref{subsubsec:Xcube-fracton}.  We need to show that for $e \in {\cal L}_b$, $ \pi(e) = 0$ implies $e \in {\cal L}_b$.  That is we need to show that any excitation with trivial plane charges is locally createable. Our strategy will be to start with an arbitrary such excitation, and reduce it to the trivial configuration by acting with a finite product of $Z_\ell$ Pauli operators. Among those planes of constant $z$ that contain at least one cube excitation, we find the plane with $z = z_0$ such that no cube excitations are present for $z < z_0$.
 By acting with $Z_\ell$ on a suitable link $\ell$, we can create two fractons on nearest-neighbor cubes in this plane, while also creating two fractons in the $z = z_0 + 1$ plane. This allows us to annihilate any pair of cube excitations in the $z = z_0$ plane, independent of their location, at the expense of creating some excitations in the $z = z_0 + 1$ plane.  Since by assumption the $z = z_0$ plane contains an even number of cube excitations, we can eliminate all excitations in this plane.  This process can be repeated to eliminate all cube excitations except those in a single  $z = z_1$ plane.
 
The next step is to divide the $z = z_1$ plane into columns with fixed $x$-coordinate.  Among those columns containing at least one cube excitation, we find the $x = x_0$ column, where $x_0$ is such that no cube excitations are present for $x < x_0$.  This column must contain an even number of cube excitations, because the $yz$-plane containing it contains an even number of cube excitations.  We can then act with $Z_\ell$ on suitable links to move pairs of cube excitations from the $x = x_0$ column to the $x = x_0 + 1$ column, until no cube excitations with $x = x_0$ remain.  We repeat this process until all excitations reside in a single column with $x = x_1$ (and $z = z_1$).  At this point, there can be no cube excitations remaining, because when all the cube excitations reside in a single column, the only way for all plane charges to be trivial is to have no excitations at all.  This shows that any excitation with trivial plane charges is locally createable.

\begin{figure}[b]
  \includegraphics[scale=0.8]{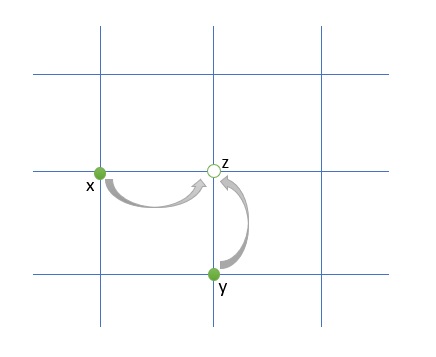}
 \caption{Reducing a configuration of $1d$ particles with trivial plane charges to that with only $s^z$ excitations.}
 \label{fig:Xcube1dhop}
 \end{figure}

Second, we show that $p \in {\cal P}_b$ if and only if the integers $k_p, l_p, m_p$ are either all even or all odd.  It is obvious that any sum of generators $P_{yz}(x) + P_{xz}(y) + P_{xy}(z)$ of ${\cal P}_b$  gives an element with $k_p,l_p,m_p$  all even or all odd. So we just have to show that any element $p \in {\cal P}$ with $k_p, l_p, m_p$ either all even or all odd can be obtained as a sum of such generators.
Equivalently, we can add generators to the element in question to reduce it to the identity.  First, we can obviously add generators to produce a new element where $m_{p} = 0$, and where $k_{p}$ and $l_{p}$ are even.  Now we add generators to set $l_{p} = 0$ without changing $m_{p}$ or $k_p$.  To do this, let $y_1$ and $y_2$ be two values of $y$ for which $Q_{xz}(y) = 1$; there must be at least two such values because $l_{p}$ is even.  We then add the two generators $P_{yz}(x) + P_{xz}(y_1) + P_{xy}(z)$ and $P_{yz}(x) + P_{xz}(y_2) + P_{xy}(z)$, where $x$ and $z$ are arbitrary.  The $P_{yz}$ and $P_{xy}$ terms cancel, and the $P_{xz}$ terms set $Q_{xz}(y_1) = Q_{xz}(y_2) = 0$.  We can continue this procedure until $l_p = 0$.  The same procedure can then be applied to set $k_p = 0$ without changing $l_p$ or $m_p$, and we thus reduce $p$ to zero by adding generators.

\subsection{Lineon fusion theory of the X-cube model}
\label{app:Xcube-lineon}

First we show that ${\cal L}_a = \operatorname{ker} \pi$.  First, it is obvious that ${\cal L}_a \subset \operatorname{ker} \pi$, because the generators map to $0$ under $\pi$.  To show $\operatorname{ker} \pi \subset {\cal L}_a$, and thus ${\cal L}_a = \operatorname{ker} \pi$, consider an excitation $e \in \operatorname{ker} \pi$.  Our first step will be to eliminate all the $\mu = x,y$ vertex excitations in favor of $\mu = z$ vertex excitations, by adding generators of ${\cal L}_a$ to $e$.  Expressing $e$ using Eq.~(\ref{eqn:general-lineon}), we have
\begin{eqnarray}
0 = \pi(e) &=& \sum_{(x,y,z)} \Big[ n_x(x,y,z) \Big( P_{xy}(z) + P_{xz}(y) \Big) \nonumber  \\
&+& n_y(x,y,z) \Big( P_{xy}(z) + P_{yz}(x) \Big)  \\
&+& n_z(x,y,z) \Big( P_{xz}(y) + P_{yz}(x) \Big)  \Big] \text{.} \nonumber
\end{eqnarray}
Focusing on the $z = z_0$ plane, we observe that the coefficient of $P_{xy}(z_0)$ in this expression must be even, and is equal to $\sum_{x,y} ( n_x(x,y,z_0) + n_y(x,y,z_0) )$.  That is, the total number of $\mu = x,y$ vertex excitations in the $z = z_0$ plane is even.  Moreover, because \emph{e.g.} a $\mu = x$ excitation is equivalent to a $\mu = y$ and $\mu = z$ excitation by the relation Eq.~(\ref{eqn:cage-reln}), we can take the number of $\mu = x$ and $\mu = y$ excitations to be equal.  Any pair of $\mu = x$ and $\mu = y$ excitations can moved (\emph{i.e.} by adding ${\cal L}_a$ generators) to the point where their lines of motion intersect, where they fuse together to a $\mu = z$ excitation, as illustrated in Fig.~\ref{fig:Xcube1dhop}.  Repeating this procedure in every $xy$-plane, we obtain a new configuration with only $\mu = z$ vertex excitations.

\begin{figure}[b]
  \includegraphics[scale=0.5]{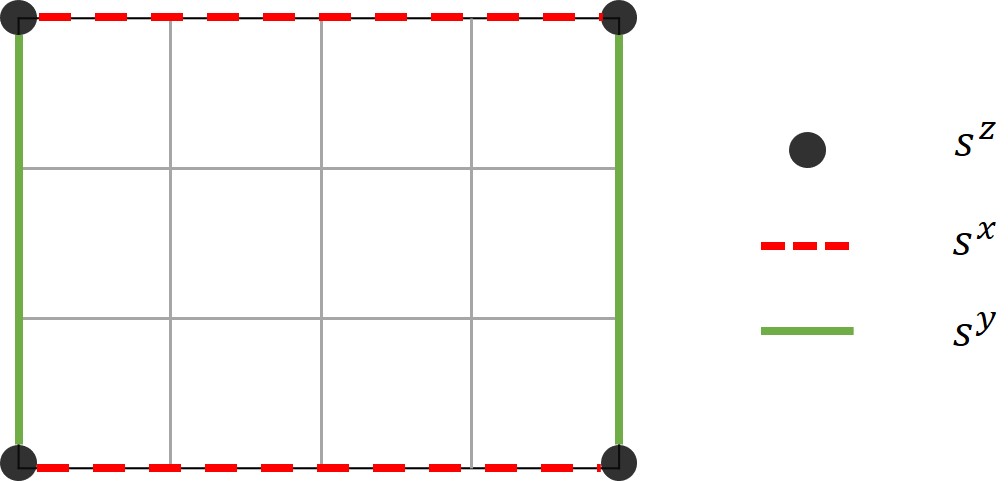}
 \caption{Local moves needed to ``clean up'' a configuration of trivial plane charges to the vacuum.}
 \label{fig:Xcube1dcleanup}
 \end{figure}

The $\mu = z$ excitations can then all be moved into the same $xy$-plane, say with $z = 0$, resulting in a new excitation $e'$.  $\pi(e') = 0$ implies that $e'$ has an even number of $\mu = z$ excitations on every row and column of the $z = 0$ plane.  By Eq.~(\ref{eqn:cage-reln}), each of these excitations is equivalent to a pair of $\mu = x$ and $\mu = y$ excitations.  In every row, we have an even number of $\mu = x$ excitations, which can be brought together and annihilated.  Similarly, the $\mu = y$ excitations can be brought together and annihilated in every column.  This process is illustrated in Fig.~\ref{fig:Xcube1dcleanup}.  It follows that ${\cal L}_a = \operatorname{ker} \pi$.

Now, we show that $p \in {\cal P}_a$ if and only if $k_p + l_p + m_p = 0 \mod 2$.  It is obvious that any sum of the generators in Eq.~(\ref{eqn:Sa-generators}) satisfies $k_p + l_p + m_p = 0 \mod 2$.  Conversely, we suppose that $p$ satisfies $k_p + l_p + m_p = 0 \mod 2$, and we will show that $p$ can be brought to the trivial element by adding generators.  Without loss of generality, assume $k_p \leq l_p \leq m_p$.  Then each of the $k_p$ $P_{yz}(x)$ terms appearing in $p$ can be grouped with a  $P_{xz}(y)$ term to form a generator that we subtract from $p$.  This eliminates the $P_{yz}(x)$ terms, and leaves $(l_p - k_p)$ $P_{xz}(y)$ terms, which we then each group with a $P_{xy}(z)$ term to form a generator.  We thus eliminate all the $P_{yz}(x)$ and $P_{xz}(y)$ terms, leaving $(m_p + k_p - l_p)$ $P_{xy}(z)$ terms, with  $(m_p + k_p - l_p)$ even.  Each of these terms can be replaced by a generator according to $P_{xy}(z) \to P_{xy}(z) + P_{yz}(x_0)$, for a fixed $x_0$, without affecting the value of $p$, because $P_{yz}(x_0)$ appears in the sum an even number of times and cancels out.  This establishes the desired result.

\subsection{Fusion theory of the checkerboard model}
\label{app:checkerboard}

We first show that $\operatorname{ker} \pi \subset {\cal L}_a$.  That is, given $e \in {\cal E}_a$ such that $\pi(e) = 0$, we would like to show that $e$ can be reduced to zero by adding generators of ${\cal L}_a$.  It is clear that $\pi(e) = 0$ implies the number of excitations in every plane of constant $z$ is even.  We first focus on the plane with $z = z_0$, where $z_0$ is chosen so that there are no excitations with $z > z_0$.  A subset of the generators of ${\cal L}_a$ can be written
\begin{widetext}
\begin{eqnarray}
&& f(x,y,z_0) + f(x-1,y+1,z_0)  + f(x-1,y,z_0 - 1) + f(x,y-1,z_0-1) \text{,} \nonumber \\
&& f(x,y,z_0) + f(x+1,y+1,z_0)  + f(x,y-1,z_0-1) + f(x+1,y,z_0-1)  \text{,} \nonumber
\end{eqnarray}
\end{widetext}
where $x$ and $y$ are arbitrary as long as $(x,y,z_0) \in \Lambda$.  Ignoring the $z = z_0 - 1$ terms, these generators correspond to arbitrary nearest-neighbor pairs of excitations in the $z = z_0$ plane.  Since the total number of excitations in this plane is even, we can add such generators to remove all $z = z_0$ excitations, at the possible expense of creating some new excitations with $z = z_0 -1$. 

We can  repeat the above procedure, with decreasing $z_0$, until we obtain a new configuration $e' \in {\cal E}_a$, such that all excitations are contained in a single plane of fixed $z = z_1$, and still $\pi(e') = 0$.  It is possible to sum generators of ${\cal L}_a$ to obtain elements of ${\cal L}_a$ with support only in the $z = z_1$ plane, of the form
\begin{widetext}
\begin{equation}
f(x,y,z_1) + f(x+2,y,z_1) + f(x,y+2,z_1) + f(x+2, y+2,z_1) \text{,} \label{eqn:z1planeelements}
\end{equation}
\end{widetext}
where $x$ and $y$ are arbitrary such that $(x,y,z_1) \in \Lambda$.  $\pi(e') = 0$ implies that there are an even number of excitations in every row and column in the $z = z_1$ plane, and we can add elements of the form in Eq.~(\ref{eqn:z1planeelements}) to eliminate all excitations except in two adjacent columns, in a new configuration $e''$.  At this point, there is only a single cube in each row that could host an excitation, so the constraint of an even number of excitations in each row implies $e'' = 0$, and we have finished showing $\operatorname{ker} \pi \subset {\cal L}_a$

Now we show that $\tilde{\cal P} = {\cal P}_a$.  First we show that ${\cal P}_a \subset \tilde{\cal P}$.  If $p \in {\cal P}_a$, then $p = \pi(e)$ for some $e \in {\cal E}_a$.  Writing $e = \sum_{\br \in \Lambda} n(\br) f(\br)$ with $n(\br) \in \{0 ,1\}$, we define quantities
\begin{eqnarray}
N_{eee} &=& \sum_{x,y,z \,\, \text{even}} n(x,y,z) \mod 2 \\
N_{eoo} &=& \sum_{x \,\, \text{even}} \sum_{y,z \,\, \text{odd}} n(x,y,z) \mod 2 \\
N_{oeo} &=& \sum_{y \,\, \text{even}} \sum_{x,z \,\, \text{odd}} n(x,y,z) \mod 2 \\
N_{ooe} &=& \sum_{z \,\, \text{even}} \sum_{x,y \,\, \text{odd}} n(x,y,z) \mod 2 \text{.}
\end{eqnarray}
It is straightforward to express $k_p^{o,e}$, $l_p^{o,e}$ and $m_p^{o,e}$ in terms of these quantities by
\begin{eqnarray}
k^o_p \mod 2 &=& N_{oeo} + N_{ooe}  \mod 2  \\
l^o_p  \mod 2 &=& N_{eoo} + N_{ooe}  \mod 2 \\
m^o_p \mod 2 &=& N_{eoo} + N_{oeo} \mod 2  \\
k^e_p  \mod 2 &=& N_{eee} + N_{eoo}   \mod 2 \\
l^e_p \mod 2 &=& N_{eee} + N_{oeo}   \mod 2\\
m^e_p \mod 2 &=& N_{eee} + N_{ooe}  \mod 2 \text{.}
\end{eqnarray}
From these expressions it is straightforward to check that Eqs.~(\ref{eqn:check1}-\ref{eqn:check4}) hold, which shows that $p \in \tilde{\cal P}$ and so ${\cal P}_a \subset \tilde{\cal P}$.

Next we show $\tilde{\cal P} \subset {\cal P}_a$.  That is, given $p \in \tilde{\cal P}$, we need to show there exists $e \in {\cal E}_a$ such that $p = \pi(e)$.  Equivalently, we can start with a general $p \in \tilde{\cal P}$ and add elements of the form $\pi(e)$ until $p$ is reduced to zero.  First, it is clearly possible to add elements $\pi(e)$ to set $Q_{xy}(z) = 0$ for all $z$, resulting in a new element $p$ with $m^o_p = m^e_p = 0$.  This results in simplified constraints
\begin{eqnarray}
k^o_p + l^o_p  &=& 0 \mod 2  \\
k^o_p + l^e_p  &=& 0 \mod 2 \\
k^e_p + l^o_p &=& 0 \mod 2 \\
k^e_p + l^e_p &=& 0 \mod 2   \text{.}
\end{eqnarray}
These constraints are solved by either all $k^{o,e}_p = l^{o,e}_p = 0 \mod 2$, or all $k^{o,e}_p = l^{o,e}_p = 1 \mod 2$.  

Now suppose that $Q_{xz}(y) = 1$ for some $y$, then there must be some $y' \neq y$ for which also $Q_{xz}(y') = 1$, in order to satisfy the constraints.  We can then add $\pi[ f(x,y,z) + f(x',y',z) ]$ to set $Q_{xz}(y) = Q_{xz}(y') = 0$,while keeping all $Q_{xy}(z) = 0$, and this can be repeated until $Q_{xz}(y) = 0$ for all $y$.  

This results in a new element $p$ where ${k^o_p = k^e_p = 0 \mod 2}$ and $Q_{xz}(y) = Q_{xy}(z) = 0$ for all $y$ and all $z$.  Suppose $Q_{yz}(x) = 1$, then there must be $x' \neq x$, but with $x'$ of the same parity as $x$, where also $Q_{yz}(x')  = 1$.  We can then add $\pi[ f(x,y,z) + f(x',y,z) ]$ to set $Q_{yz}(x) = Q_{yz}(x') = 1$ without changing any of the $Q_{xz}(y)$ or $Q_{xy}(z)$.  This can then be repeated until $p$ is set to zero, which shows $\tilde{\cal P} \subset {\cal P}_a$ as desired.

\subsection{Fusion theory for electric charges in the $(1,1)$ scalar charge theory}
\label{app:11electric}

We would like to show that $\pi_S : {\cal E}_e / {\cal L}_e \to {\cal S}_e$ is an isomorphism.  First we will show that $\operatorname{ker} \pi = {\cal L}_e$, which implies $\pi_S$ is injective.  In Sec.~\ref{sec:11scalarcharge} we pointed out that $ {\cal L}_e \subset \operatorname{ker} \pi$, so here we just need to show $\operatorname{ker} \pi \subset {\cal L}_e$.  Given $e \in {\cal E}$ with $\pi(e) = 0$, we will reduce $e$ to zero by adding generators of ${\cal L}_e$.  By adding generators of the form $2 f(\br) - f(\br - \hz) - f(\br + \hz)$ (and those related to it by cubic rotations), \emph{any} excitation $e \in {\cal E}_e$ can be reduced to one with $q(\br) \neq 0$ only within a $2 \times 2 \times 2$ cube, which we take to have opposite corners at $\br = 0$ and $\br = \hx + \hy + \hz$, so we take $e$ to be of this form.

Now, $\pi(e) = 0$ implies the total charge and dipole moment of the charge configuration on the cube both vanish.  Consider two opposing square faces of the cube, which for convenience we call the top and bottom faces, and let $q_{{\rm top}}$ be the charge on the top face.  Then the charge on the bottom face is $-q_{{\rm top}}$ because the total charge vanishes.  But because the dipole moment also must vanish, we have $q_{{\rm top}} = 0$.  Therefore the total charge on all the square faces of the cube is zero.

\begin{figure}
  \includegraphics[scale=0.7]{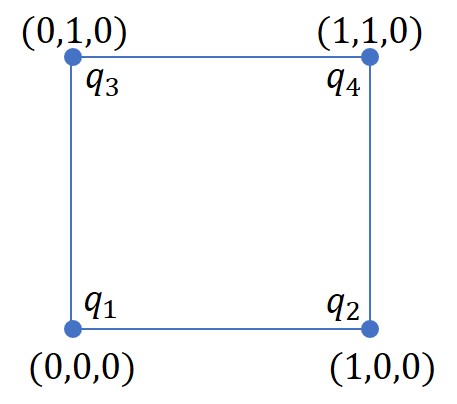}
 \caption{The $z=0$ face of the $2 \times 2 \times 2$ cube with zero charge and dipole moment.  The four vertices have charges $q_1, \dots, q_4$ as shown, and positions $\br_1 = (0,0,0)$, $\br_2 = (1,0,0)$, and so on.}
 \label{fig:sctcleanup}
 \end{figure}

Next consider the $z = 0$ face of the cube as illustrated in Fig.~\ref{fig:sctcleanup}.  The charges $q_1, \dots, q_4$ at the corners satisfy 
\begin{equation}
\sum_{i=1}^{4} q_{i} =0,\ \sum_{i=1}^{4}x_{i}q_{i}=0,\ \sum_{i=1}^{4}y_{i}q_{i}=0  \text{.}
\end{equation}
The solution to these equations is $q_1 = q_4 = -q_2 = -q_3 = n$, and this charge configuration is $n$ times the generator $f(0) - f(\hx) - f(\hy) + f(\hx + \hy)$, so we can set all the charges on this face to zero by adding generators.  The same holds for the $z=1$ face, so we can reduce $e$ to zero by adding ${\cal L}_e$ generators.  Therefore we have shown that $\pi_S$ is injective.

The next step is to show that $\pi_S$ is surjective.  Let
\begin{equation}
e = Q f(0) + D_x [ f(\hx) - f(0) ] + D_y [ f(\hy) - f(0) ] + D_z [ f(\hz) - f(0) ] \text{,}
\end{equation}
for $Q, D_x, D_y, D_z$ arbitrary integers.  Then clearly 
\begin{equation}
\pi(e) = Q \alpha + \boldsymbol{D} \cdot \boldsymbol{\beta}  \text{.}
\end{equation}

\subsection{Fusion theory for magnetic charges in the $(1,1)$ scalar charge theory}
\label{app:11magnetic}

Here we show that $\pi_S : {\cal E}_m / {\cal L}_m \to {\cal S}_m$ is an isomorphism.  It is easily checked that ${\cal L}_m \subset \operatorname{ker} \pi$, so to show $\pi_S$ is injective we need to establish $\operatorname{ker} \pi \subset {\cal L}_m$.   First, any element $e \in {\cal E}_m$ can be reduced to one of the form
\begin{equation}
e = \sum_{n \in \z, \mu}  q(n \hat{\mu}, \mu) g(n \hat{\mu}, \mu)  \text{,}
\end{equation}
where $\hat{\mu} = \hx, \hy, \hz$, by adding ${\cal L}_m$ generators of the form shown in Fig.~\ref{fig:Lm_generators}(a).  Moreover, by adding ${\cal L}_m$ elements of the form illustrated in Fig.~\ref{fig:Lm_generators}(c), $e$ can be further reduced so that $q(n \hat{\mu}, \mu) = 0$ unless $n=0, -1$.  Now we use $\pi(e) = 0$.  The vanishing of the coefficient of $\alpha_{\mu}$ in $\pi(e)$ implies 
\begin{equation}
g(0,\mu) + g(-\hat{\mu}, \mu) = 0 \text{.}
\end{equation}
Therefore we have
\begin{equation}
e = \sum_{\mu} n_{\mu} [ g(0,\mu) - g(-\hat{\mu}, \mu) ] \text{,}
\end{equation}
for integers $n_{\mu}$.  Then the vanishing of the coefficient of $\beta$ implies $\sum_{\mu} n_{\mu} = 0$.  Therefore,
\begin{eqnarray}
e &=& n_x [ g(0,x) - g(-\hx, x) - g(0,z) + g(-\hz,z)]  \\ 
&+& n_y [ g(0,y) - g(-\hy, y) - g(0,z) + g(-\hz,z)] \text{,}   \nonumber
\end{eqnarray}
which is a linear combination of ${\cal L}_m$ generators,  so $e \in {\cal L}_m$.

Next, to show that $\pi_S$ is surjective, we consider $e \in {\cal E}_m$ with
\begin{equation}
e = Q_x g(0,x) + Q_y g(0,y) + Q_z g(0,z) + P [ g(0,x) - g(-\hx, x) ] \text{,}
\end{equation}
where $Q_{\mu}$ and $P$ are arbitrary integers.  We have
\begin{equation}
\pi(e) = \boldsymbol{Q} \cdot \boldsymbol{\alpha} + P \beta \text{,}
\end{equation}
so $\pi$ and $\pi_S$ are clearly surjective.

\bibliography{ref}

\end{document}